\documentclass[apj,numberedappendix,appendixfloats]{emulateapj}
\usepackage{amsmath}
\usepackage{natbib}
\slugcomment{}        

\shorttitle{Local Analogs}
\shortauthors{F. Bian et al.}

\begin{document}


\title{Local Analogs for High-redshift Galaxies: Resembling the Physical Conditions of the
Interstellar Medium in High-redshift Galaxies.}


\author{Fuyan Bian\altaffilmark{1,4}, Lisa J. Kewley\altaffilmark{1}, Michael A. Dopita\altaffilmark{1,2}, Stephanie Juneau\altaffilmark{3}}
\altaffiltext{1}{Research School of Astronomy \& Astrophysics, Mt Stromlo Observatory, Australian National University, Weston Creek, ACT, 2611, Australia}
\altaffiltext{2}{Department of Astronomy, King Abdulaziz University, P.O. Box 80203 Jeddah, Saudi Arabia}
\altaffiltext{3}{CEA-Saclay, DSM/IRFU/SAp, 91191 Gif-sur-Yvette, France}
\altaffiltext{4}{Stromlo Fellow}



\begin{abstract}
We present a sample of local analogs for high-redshift galaxies selected in the Sloan Digital Sky Survey (SDSS). The physical conditions of the interstellar medium (ISM) in these local analogs resemble those in high-redshift galaxies. These galaxies are selected based on their positions in the [\ion{O}{3}]/H$\beta$ versus [\ion{N}{2}]/H$\alpha$ nebular emission-line diagnostic diagram. We show that these local analogs share similar physical properties with high-redshift galaxies, including high specific star formation rates (sSFRs), flat UV continuums and compact galaxy sizes. In particular, the ionization parameters and electron densities in these analogs are comparable to those in $z\simeq2-3$ galaxies, but higher than those in normal SDSS galaxies by $\simeq$0.6~dex and $\simeq$0.9~dex, respectively. The mass-metallicity relation (MZR) in these local analogs shows $-0.2$~dex offset from that in SDSS star-forming galaxies at the low mass end, which is consistent with the MZR of the $z\sim2-3$ galaxies. We compare the local analogs in this study with those in other studies, including Lyman break analogs (LBA) and green pea (GP) galaxies. The analogs in this study share a similar star formation surface density with LBAs, but the ionization parameters and electron densities in our analogs are higher than those in LBAs by factors of 1.5 and 3, respectively. The analogs in this study have comparable ionization parameter and electron density to the GP galaxies, but our method can select galaxies in a wider redshift range. We find the high sSFR and SFR surface density can increase the electron density and ionization parameters, but still cannot fully explain the difference in ISM condition between nearby galaxies and the local analogs/high-redshift galaxies. 

\end{abstract}




\section{Introduction}
In the last decade, our knowledge of high-redshift galaxies has improved tremendously by studying a large sample of star-forming galaxies beyond $z\sim1$ \citep[e.g.,][]{Steidel:2003kx,Steidel:2004fk,Franx:2003fk,Daddi:2004fk,Daddi:2005fk,Bouwens:2007lr,Hu:2010pd,Stark:2010lr}. Studies on these high-redshift galaxies have suggested that high-redshift star-forming galaxies have different properties compared to local star-forming galaxies. High-redshift star-forming galaxies are characterized by higher specific SFRs by about 1~dex \citep[sSFR$ = $SFRs/$M_{\ast}$, e.g.,][]{Daddi:2007qy,Gonzalez:2010fk,Reddy:2012aa,Stark:2013aa}, smaller galaxy sizes by 0.3-0.5~dex \citep[e.g.,][]{Ferguson:2004lr,Trujillo:2006aa,Franx:2008aa,Mosleh:2012vn,van-der-Wel:2014aa}, higher gas fractions by a factor of $\sim5$ \citep[e.g.,][]{Daddi:2010fk,Tacconi:2010qy,Genzel:2010aa,Genzel:2013aa,Tacconi:2013aa,Saintonge:2013aa}, and clumpy thick star-forming disks with higher velocity dispersion \citep[e.g.,][]{Forster-Schreiber:2009cr,Forster-Schreiber:2011qy,Genzel:2011aa,Newman:2013aa}. 

Studies also suggest that the interstellar medium (ISM) conditions in high-redshift galaxies are different from those in nearby galaxies. Observationally, there exists an offset between the high-redshift star-forming galaxy sequence and the low-redshift star-forming galaxy sequence in the [\ion{O}{3}]$\lambda$5007/H$\beta$ versus [\ion{N}{2}]$\lambda$6583/H$\alpha$ ``Baldwin-Phillips-Terlevich'' \citep[BPT,][]{Baldwin:1981rr} diagram. \citep[e.g.,][]{Liu:2008kx,Brinchmann:2008ab,Overzier:2008aa,Overzier:2009aa,Hainline:2009aa,Bian:2010vn,Steidel:2014aa}. Recently, \citet{Steidel:2014aa} have found a well-defined high-redshift locus in the BPT diagram with a large sample (126) of $z\simeq2.3$ UV-selected star-forming galaxies for the first time. A number of studies have suggested that the offset between the high-redshift and local BPT locus could be due to higher ionization parameters, harder ionization radiation fields, various N/O ratios, AGN/shock contributions, and/or selection effects \citep[e.g.,][]{Kewley:2013ab,Kewley:2013aa,Juneau:2014aa,Masters:2014aa,Newman:2014aa,Steidel:2014aa,Shapley:2015aa}.

The dramatic change in ISM conditions between low-redshift and high-redshift galaxies raises the following crucial question. What are the major causes of the change in ISM conditions? Are the local empirical metallicity calibrations based on strong emission-line diagnostics and $T_e$-based metallicities still applicable to high-redshift star-forming galaxies? Unfortunately, it is difficult to fully probe the ISM condition of high-redshift galaxies due to their low surface brightness and small angular size. An alternative method to approach this problem is to select a sample of local analogs whose ISM properties mimic those in high-redshift galaxies. Using these local analogs, we can better understand the ISM conditions in high-redshift galaxies and guide the metallicity calibrations in high-redshift galaxies.

A variety of methods have been developed to select local analogs for high-redshift galaxies. Most of the methods are based on the global properties of galaxies, such as their far-UV (FUV) surface luminosity \citep[e.g., Lyman break analogs (LBA)][]{Heckman:2005aa,Hoopes:2007aa,Stanway:2014aa}, H$\alpha$ luminosity \citep[e.g.,][]{Green:2014aa}, or extremely strong [\ion{O}{3}]$\lambda$5007 emission of the equivalent width (EW) up to 1000~{\AA} \citep[e.g., green pea galaxies,][]{Cardamone:2009aa}, or both H$\alpha$ and [\ion{O}{3}] luminosities \citep[e.g.,][]{Juneau:2014aa}. These galaxies share many similar physical properties with high-redshift star-forming galaxies, including the properties of metallicity, dust extinction, morphology, outflow, kinematics, and molecular gas \citep[e.g.,][]{Basu-Zych:2007aa, Basu-Zych:2009aa, Overzier:2008aa,Overzier:2009aa,Overzier:2010aa, Overzier:2011aa,Goncalves:2010aa,Goncalves:2014aa, Heckman:2011aa, Green:2014aa,Bassett:2014aa,Fisher:2014aa}. On the other hand, \citet{Shirazi:2014ab} pointed out that the low-redshift galaxies with similar global properties, e.g., stellar mass, SFR, and sSFR, do not necessarily to reproduce the similar ISM conditions in high-redshift galaxies.

It is therefore essential to select a sample of local analog galaxies that share the same ISM conditions as high-redshift star-forming galaxies. This type of local analog for high-redshift galaxies could provide us with better understanding of the cause of the different ISM conditions and allow us to improve the metallicity measurements in high-redshift star-forming galaxies. \citet{Steidel:2014aa} suggested that the so-called 'green pea' (GP) galaxies, which are also known as 'isolated extragalactic HII region galaxies' \citep{Sargent:1970aa},  are probably good local analogs for high-redshift galaxies in terms of ISM conditions, and they are located in the high-redshift star-forming locus in the BPT diagram. The GP galaxies are selected based on their distinctive colors due to the very strong optical emission lines \citep[EW$_0>100$~{\AA},][]{Cardamone:2009aa}. In this study, we select galaxies that are located in the high-redshift BPT locus as defined in \citet{Steidel:2014aa} without applying any prior selections on the emission line strength. We demonstrate that the galaxies selected by our method share the same properties, especially the physical conditions of the ISM, as star-forming galaxies at $z\sim2-3$. Therefore, this type of analog could provide us with a local laboratory to study the cause of extreme star formation and ISM condition in the high-redshift galaxies.

The paper is organized as follows. In section~\ref{sec:selection}, we describe the selection method of local analogs for high-redshift galaxies. In Section~\ref{sec:property}, we compare the properties in the local analogs with those in both nearby SDSS and high-redshift star-forming galaxies. In Section~\ref{sec:other_analog}, we compare the local analogs in this study with those in other studies, including the LBAs, GP galaxies and SDSS galaxies with high line luminosities. In Section~\ref{sec:discussion}, we discuss whether or not the high sSFR and SFR surface density can fully explain the ISM evolution from normal nearby star-forming galaxies to the local analogs and high-redshift galaxies. Throughout the paper, we use the value of the SFR and stellar mass based on the \citet{Chabrier:2003wd} initial mass function (IMF). The SFRs and stellar masses derived from a \citet{Salpeter:1955kx} or a \citet{Kroupa:2001vn} IMF were normalized to the Chabrier IMF, by multiplying the results by 0.55 or 0.89, respectively. We use the following cosmological parameters for calculations: Hubble constant, $H_0=70$~km~s$^{-1}$~Mpc$^{-1}$; dark matter density, $\Omega_{\rm M} = 0.30$; dark energy density, and $\Omega_{\rm M} = 0.70$ for a flat Universe \citep[e.g.,][]{Spergel:2007fk}.


\begin{figure}[htbp]
\begin{center}
\includegraphics[scale=0.33,angle=-90]{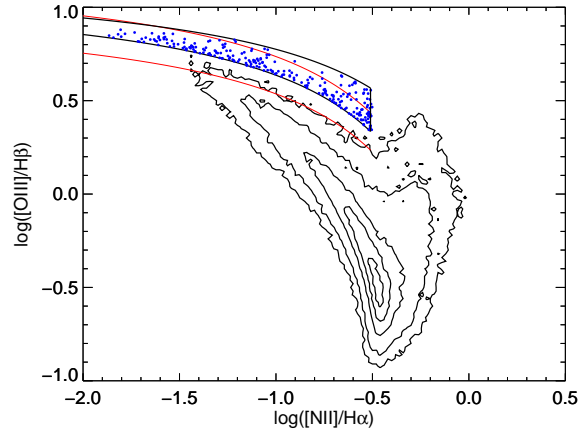}
\caption{BPT diagram of local star-forming galaxies. The contours represent the distribution of local star-forming galaxies in the BPT diagram and region bounded by the black solid lines is the selection criteria for our local analogs. The region between the two red lines is the high-redshift star-forming locus in the BPT diagram \citep{Steidel:2014aa}. The blue filled points are the local analogs selected in study.\label{fig_select}}
\end{center}
\end{figure}
\section{The Selection of Local Analogs for High-redshift Galaxies} \label{sec:selection}
We select our local analogs from a sample of nearby galaxies ($z<0.35$) observed by the Sloan Digital Sky Survey \citep[SDSS;]{York:2000aa} by using the MPA-JHU value added catalog for SDSS Data Release 7 \citep[DR7,][]{Abazajian:2009aa}. This catalog includes 819,333 unique galaxies and provides information on the flux, EW, and line width of the rest-frame optical nebular emission lines (e.g., [\ion{O}{2}]$\lambda3727$, H$\beta$, [\ion{O}{3}]$\lambda\lambda4959, 5007$, H$\alpha$, [NII]$\lambda6583$, and [\ion{S}{2}]$\lambda\lambda6717,6731$), stellar masses \citep[$M_*$,][]{Kauffmann:2003aa}, star formation rates \citep[SFRs,][]{Brinchmann:2004fk}, and specific SFRs \citep[SFR/$M_{*}$, sSFRs,][]{Brinchmann:2004fk}. In this study, we will adopt the emission-line and SFR measurements from the MPA-JHU catalogs, and measure the stellar mass with our own procedures; see details in section~\ref{sec:mass}.

\subsection{Selection Criteria}
\citet{Brinchmann:2008ab} and \citet{Liu:2008kx} investigated the relation between the ISM conditions and their locations in the BPT diagram, and found that local galaxies with high EW(H$\alpha$) and high electron density lie well above the ridge line of the star-forming locus in the BPT diagram. In this work, we will use the well-defined high-redshift star-forming sequence in the BPT diagram in \citet{Steidel:2014aa} as a guide to define a selection region for local analogs in the BPT diagram.

We select the local analogs for high-redshift galaxies as follows: (1) We select the objects that are classified as star-forming/starburst galaxies in the MPA/JHU catalog, in which  \citet{Kewley:2001aa} criteria were used to separate the star-forming/starburst galaxies from AGNs; (2) We select galaxies with signal-to-noise ratios (S/N) of
[\ion{O}{2}]$\lambda$3727, H$\beta$, [\ion{O}{3}]$\lambda\lambda4959, 5007$, H$\alpha$ and [\ion{N}{2}]$\lambda$6583 emission lines greater than 10; (3) We apply the following criteria to select galaxies that shares the similar BPT locations of high-redshift star-forming galaxies:
\begin{equation}
\log({\rm[OIII]/H\beta)})>\frac{0.66}{\log({\rm[NII]/H\alpha)})-0.31}+1.14 \label{bottom},
\end{equation}
\begin{equation}
\log({\rm[OIII]/H\beta)})<\frac{0.61}{\log({\rm[NII]/H\alpha)})-0.47}+1.19 \label{top},
\end{equation} 
and
\begin{equation}
\log({\rm[NII]/H\alpha)}) < -0.5\label{right}
\end{equation}
Figure~\ref{fig_select} shows the selection region of local analogs for high-redshift galaxies bounded by the black solid lines. The region between the two red lines represents the high-redshift star-forming locus in the BPT diagram from \citet{Steidel:2014aa}. The bottom boundary of the selection criteria (equation~\ref{bottom}) is about 0.2~dex higher than that in the high-redshift locus to reduce contaminations from the normal nearby star-forming galaxies. We adopt the top boundary (equation~\ref{top}) from equation 5 in \citet{Kewley:2001aa}, which is defined as the maximum starburst model. The right boundary (equation~\ref{right}) is used to reduce the contamination from AGN, and this criterion also tends to select low-metallicity galaxies. The selection criteria overlap with the high-redshift star-forming BPT locus and are well separated from the local BPT locus as well as the well-known AGN regions to reduce contamination. (4) We visually inspected the images of these galaxies and their fiber positions to remove spurious galaxies being targeted due to poor photometric deblending \citep{Andrews:2013aa}. 
 A total of 252 galaxies satisfy the above selection criteria (blue filled circles in Figure~\ref{fig_select}). The redshift distribution of the galaxies is $z = 0.05-0.35$ with a median redshift of $z=0.1$. In this redshift range, the galaxy angular size is small enough so that the fiber aperture effects are negligible \citep{Kewley:2005aa}. Though we apply this conservative selection to galaxies with high S/N (S/N$>$10) spectra, there may exist a fraction of contaminants (see details in section~\ref{sec:sSFR})

\subsection{Stellar Mass}\label{sec:mass}
The stellar mass measurement in the MPA-JHU catalog does not take into account the emission-line contribution to the broadband photometry. This approach is suitable for the majority of the SDSS galaxies whose emission-line equivalent widths (EWs) are $<100${\AA}. However, a large fraction ($>10$\%) of the broadband flux can come from the emission lines for galaxies with strong emission lines (EW$>100${\AA}). In this case, one would overestimate the stellar mass from fitting the broadband photometry, if the emission line contribution is not carefully removed. The local analogs selected in this work usually show strong [\ion{O}{3}]$\lambda\lambda$4959,5007 and H$\alpha$ lines with EW$>100${\AA}. Therefore, we independently derive the stellar mass rather than adopting the stellar mass estimates from MPA-JHU catalog.

We apply the IDL-based code Fitting and Assessment of Synthetic Templates \citep[FAST,][]{Kriek:2009fk} to derive the stellar mass by fitting broadband photometry. The stellar mass is calculated from the best-fit stellar synthesis model and the stellar mass uncertainty is derived using Monte Carlo simulations \citep{Kriek:2009fk}. Prior to the fitting, we correct the broadband photometry for contributions from H$\alpha$, H$\beta$, and [\ion{O}{3}] emission lines using the line EWs and redshift from the MPA-JHU catalog. We adopt the \citet[BC03]{Bruzual:2003lr} stellar synthesis models with a \citet{Chabrier:2003wd} IMF. We assume galaxies with exponentially decreasing star formation history ($\rm{SFR}\propto e^{-(t/\tau)}$) with $\log(\tau$(yr)) = 7, 8, 9, 10, and 11 and a stellar population age in the range $0-13$~Gyr. A smaller $\tau$ value corresponds to a more bursty star-formation history, and a larger $\tau$ value corresponds to a more steady star-formation history. We apply the dust extinction law proposed by \citet[][]{Calzetti:2000vn} with $E(B-V) = 0-0.6$. To investigate the contribution of the emission lines to the stellar mass measurement, we also carry out the SED fitting on the broad band photometry without correcting the emission line contribution, and we find that the broad band contribution to the stellar mass estimation is small ($<0.05$~dex) for majority of the SDSS galaxies. However, we will overestimate the stellar mass in the local analogs by a median of 0.25~dex, if we do not remove the strong emission line contributions in the local analogs.

\begin{figure}[htbp]
\begin{center}
\includegraphics[scale=0.45]{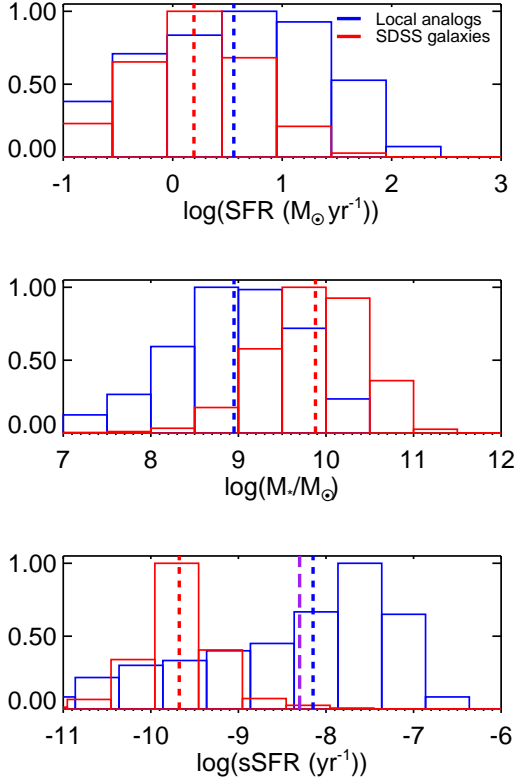}
\caption{Distribution of SFR (top panel), stellar mass (middle panel), and sSFR (bottom panel) in the local analogs for high-redshift galaxies (blue histogram) and the SDSS galaxies (red histogram). The blue and red dashed lines represent the median values of SFR, stellar mass, sSFR in the local analogs and the SDSS galaxies, respectively. The purple line in the sSFR pane represents the median value of sSFR in the high-redshift galaxies.}
\label{property}
\end{center}
\end{figure}

\section{Properties of the Local Analogs for High-redshift Galaxies}\label{sec:property}
In this section, we study the properties, including sSFR, size-mass relations, UV-continuum slope, dust attenuation properties, electron densities, and ionization parameters, of the local analogs for high-redshift galaxies. We compare these properties with those in normal SDSS galaxies and high-redshift galaxies to demonstrate that our local analogs share similar properties with high-redshift galaxies.

\subsection{Specific Star Formation Rate}\label{sec:sSFR}
We measure the SFR, stellar mass, and sSFR in the local analogs for high-redshift galaxies. Figure~\ref{property} shows the normalized distribution of the SFR, stellar mass, and sSFR of the local analogs (the blue histogram) and the SDSS galaxies (the red histogram). The blue and red dashed lines represent the median SFR, stellar mass, and sSFR of the local analogs and the local SDSS star-forming galaxies, respectively. The median SFR, stellar mass, and sSFR of the local analogs are 3.6~$M_{\sun}$~yr$^{-1}$, $9\times10^8$~$M_{\sun}$, and 7~Gyr$^{-1}$, respectively. 

We compare the sSFRs in the local analogs with those in high-redshift galaxies. Studies have shown that the sSFRs of $z\sim2-3$ star-forming galaxies are typically an order of magnitude higher than those in nearby star-forming galaxies \citep{Noeske:2007qy,Daddi:2007qy,Wuyts:2011fk,Rodighiero:2011fk,Reddy:2012aa,Salmi:2012qy,Tacconi:2013aa}. \citet{Reddy:2012aa} found that the median sSFR of $z\sim2-3$ UV-selected galaxies is $2.4$~Gyr$^{-1}$ based on the SFRs derived from the UV+24$\mu$m luminosities. This result is consistent with those found in active BzK galaxies with both the UV+24$\mu$m-based SFRs \citep{Daddi:2007qy} and the FIR-based SFRs \citep{Rodighiero:2011fk}. Our local analogs have an even higher median sSFR compared to star-forming galaxies at $z\sim2-3$ by a factor of 3. \citet{Rodighiero:2011fk} found that the sSFR becomes higher in the galaxies with lower stellar mass. The median stellar mass in our local analogs is about 10 times lower than that in the high-redshift galaxies in \citet{Reddy:2012aa}. We apply the relation between the SFR and stellar mass found in \citet{Rodighiero:2011fk} and find that the sSFR of high-redshift galaxies with stellar masses of $9\times10^8$~$M_{\sun}$ is 5~Gyr$^{-1}$ (purple dashed line in Figure~\ref{property}). The sSFR of local analogs for high-redshift galaxies is consistent with that of star-forming galaxies at $z\sim2-3$ with similar stellar mass.

It is worth noting that there exits an extended low sSFR tail down to $\log$[sSFR (yr$^{-1}$)] = -11.0, suggesting that there may exist some contaminants in our selection.

\begin{figure}[ht]
\begin{center}
\includegraphics[scale=0.35,angle=-90]{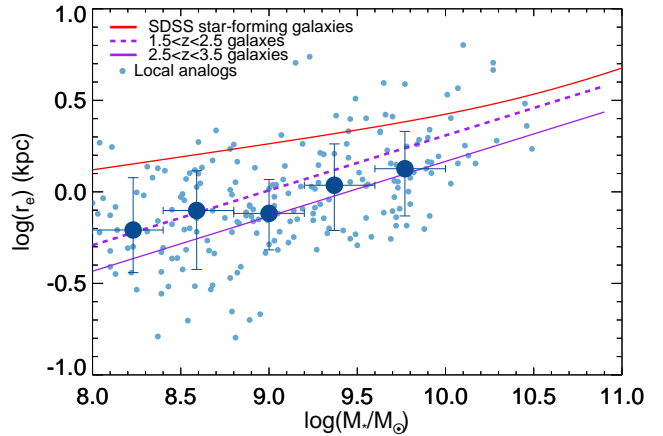}
\caption{Galaxy size and stellar mass relation. The light blue filled points represent size and mass relation in the local analogs for high-redshift galaxies. The dark blue filled points with error bars represent the median galaxy size in each mass bin. The horizontal error bars show the stellar mass range in each mass bin, and the vertical error bars represent the 16th and 84th percentiles of the galaxy  size distribution in each mass bin. The red solid line represent the galaxy size and stellar mass relation derived from SDSS star-forming galaxies \citep{Shen:2003aa}. The dashed and solid purple lines represent the galaxy size and stellar mass in UV-selected galaxies at $z\sim2$ and $z\sim3$, respectively \citep{Mosleh:2012vn}.}
\label{fig_masssize}
\end{center}
\end{figure}

\subsection{Galaxy size-mass relation}\label{size}
The relation between galaxy size and stellar mass has been established in both low- and high-redshift galaxies. Studies have found that high-redshift galaxies have smaller sizes compared to local galaxies with similar masses \citep[e.g., ][]{Trujillo:2006aa,Franx:2008aa,Mosleh:2012vn}. We compare the mass-size relation in the local analogs with those in both local star-forming galaxies and $z=2-3$ star-forming galaxies. We adopted the mass-size relation in the local star-forming galaxies established by \citet{Shen:2003aa}. The authors used the S\'ersic half-light radius in the SDSS r-band images from \citep{Blanton:2003aa} to represent the galaxy size, and they adopted stellar masses derived using the spectra features, D$_n$(4000) \& H$\delta$A \citep{Kauffmann:2003aa}, which is consistent with those derived from the broadband SED fitting.\footnote{http://www.mpa-garching.mpg.de/SDSS/DR7/mass\_comp.html}. \citet{Mosleh:2011aa} used a sample of UV-selected galaxies (BX/BM galaxies at $z\sim2$ and LBGs at $z\sim3$) to study the size-mass relation at $z\sim2$ and $z\sim3$. The galaxy sizes are measured in the deep SUBARU $K_s$-band images, and the stellar mass is derived from the SED fitting. We scale the stellar masses in \citet{Shen:2003aa} and \citet{Mosleh:2011aa}, which were derived based on a \citet{Kroupa:2001vn} IMF, to the stellar masses based on a \citet{Chabrier:2003wd} IMF. For the local analogs, we use the size measurements from \citet{Blanton:2003aa} and the stellar mass from SED fitting describe in section~\ref{sec:mass}. Figure~\ref{fig_masssize} shows the mass-size relation in the local analogs, the $z\sim2-3$ star-forming galaxies, and the SDSS star-forming galaxies. We find that the local analogs have more compact galaxy sizes than SDSS star-forming galaxies for a fixed stellar mass. The mass-size relation in local analogs is consistent with those in star-forming galaxies at $z\sim2$ and $z\sim3$. This result suggests that the local analogs have similar stellar mass surface densities to high-redshift galaxies. Considering that the sSFRs of local analogs are comparable to those of high-redshift galaxies, this relation also suggests that the the local analogs and high-redshift galaxies have similar SFR surface densities.

\begin{figure}[ht]
\begin{center}
\includegraphics[scale=0.34,angle=-90]{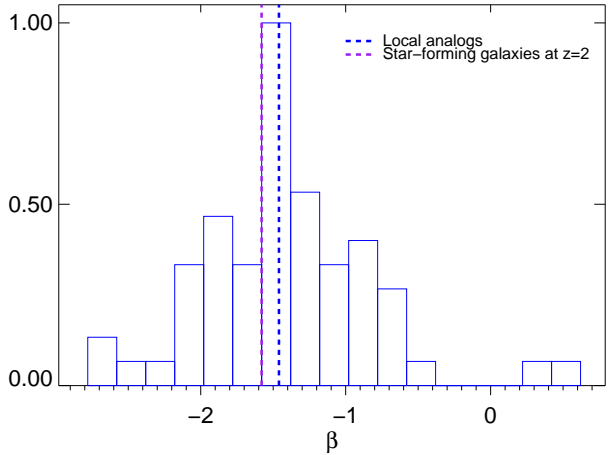}
\caption{Distribution of UV-continuum slope ($\beta$) in the local analogs for high-redshift galaxies. The blue
and purple dashed lines represent the median values of the UV-continuum slope in the local analogs and $L^*$ star-forming
galaxies at $z\sim2$ \citep{Bouwens:2009uq}.}
\label{fig_beta}
\end{center}
\end{figure}

\subsection{UV-continuum Slope}\label{sec:uv_slope}
We investigate the UV-continuum slope in our local analogs for high-redshift galaxies. The rest-frame UV continuum of galaxies can be described as a power law, $f_\lambda  \propto \lambda^\beta$. A galaxy with $\beta=-2$ corresponds to a flat spectrum in $f_\nu$, and the dust extinction leads to an increase of $\beta$. Thus, the UV-continuum slope can be used to approximate the dust obscuration in galaxies \citep[e.g.][]{Meurer:1999fk}. The average UV spectral slope, $\langle\beta\rangle$, in local starburst is around -1.3, and $\langle\beta\rangle$ is smaller (UV color bluer) at high-redshifts and lower UV luminosities \citep[e.g.,][]{Meurer:1999fk,Bouwens:2009uq,Bouwens:2012fk}. 

We measure the UV-continuum slope ($\beta$) in the local analogs for high-redshift galaxies using the Galaxy Evolution Explorer ({\it GALEX}) FUV and NUV fluxes. We cross correlate the positions of the local analogs selected in this study with the sources cataloged in GALEX-DR5 \citep{Bianchi:2011aa}. A total of 152 local analogs are detected by GALEX with $3\sigma$ in both the FUV and NUV bands. The FUV and NUV bands provide the flux at a rest-frame $\sim1400$~{\AA} and $\sim2200$~{\AA} for the galaxies at $z=0$.  At a redshift of $z>0.1$, the Ly$\alpha$ emission feature starts to shift into the FUV band and affects the UV-continuum slope measurement. \citet{Shapley:2003lr} found that the total rest-frame equivalent width of the Ly$\alpha$ emission feature is 14.3~{\AA} in the composite rest-frame UV spectrum of a sample of Lyman break galaxies (LBGs) at $z\sim3$. If we assume that our local analogs have similar Ly$\alpha$ emission strength, then the influence of the Ly$\alpha$ emission on the UV-continuum slope measurement is negligible ($\Delta\beta<0.04$).  We compute the UV-continuum slope, $\beta$, for the local analogs based on the FUV and NUV magnitudes as follows: (1) we generate a series of spectral templates with $-3<\beta<3$ with an increment of 0.01; (2) we convolve the spectral templates with the {\it GALEX} FUV and NUV band filter curves and compute the FUV and NUV band color ($\rm mag_{\rm FUV}-mag_{\rm NUV}$) for the spectral template with a given $\beta$; (3) we establish the relation between the $\beta$ and FUV$-$NUV color and fit the relation with a quadratic equation, $\beta = -2.00 + 2.32\times(\rm{mag_{FUV}-mag_{NUV}}) - 0.047\times(\rm{mag_{FUV}-mag_{NUV}})^2$; (4) we correct the {\it GALEX} magnitudes for Galactic extinction \citep{Schlegel:1998aa}; and (5) we estimate the values of $\beta$ and the uncertainties in the local analogs based on the $\rm{mag_{FUV}-mag_{NUV}}$ color. The typical errors in $\beta$ are $\sim0.51$. We calculate the weighted mean value for the UV slope using the inverse-variance weighting method and find that the weighted mean of $\beta$ is $-1.51\pm0.03$ in the local analogs.

Figure~\ref{fig_beta} shows the distribution of the $\beta$ for our local analogs and high-redshift galaxies. The blue dashed line represents the weighted mean of the UV slope for local analogs. This value is comparable to the mean UV-continuum slope, $\beta$, in $L^*$ star-forming galaxies at $z\sim2-3$ \citep[purple dashed line in Figure~\ref{fig_beta}]{Bouwens:2009uq}. We also derive their UV absolute magnitude, $M_{1400}$, which is in the range between -14 and -22, with a median value of -19.8, based on their FUV magnitudes and redshifts. This value is 1.6-2 magnitude fainter than for $L^*$ star-forming galaxies at $z\sim2-3$ \citep[e.g.,][]{Reddy:2009kx,Bian:2013aa}. The UV-continuum slope  decreases with decreasing UV luminosity. Therefore, the UV continuum in the analogs is likely to be redder than that in the $z\sim2-3$ star-forming galaxies with similar UV luminosity.

\begin{figure}[ht]
\begin{center}
\includegraphics[scale=0.36,angle=-90]{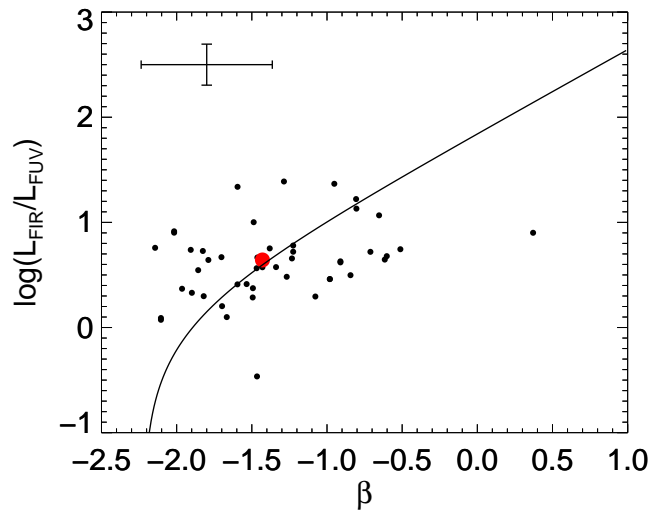}
\caption{Dust attenuation measured by the ratio between infrared and UV luminosity versus UV-continuum slope, $\beta$, in the local analogs for high-redshift galaxies. The large red circles represents the weighted mean value of dust attenuation and UV-continuum slope.  The black curve represents the relation derived form the local star-forming galaxies \citep{Meurer:1999fk}, which are also followed by typical star-forming galaxies at $z\sim2-3$. The black error bar represents the typical error of the dust attenuation and UV-continuum slope measurements in the individual galaxy.}
\label{fig_beta_fir}
\end{center}
\end{figure}

\subsection{Dust Attenuation Properties}
We investigate the relation between the dust attenuation and the UV-continuum slope in the local analogs for high-redshift galaxies and compare the relation with those in local star-forming galaxies and typical ($L^*$) star-forming galaxies at $z\sim2-3$. Studies have shown a relation between the dust attenuation, measured by the FIR-to-UV luminosity ratio ($L_{\rm FIR}/L_{\rm UV}$), and the UV-continuum slope ($\beta$) in local star-forming galaxies \citep[Meurer relation, the solid line in Figure~\ref{fig_beta_fir}, e.g.,][]{Meurer:1999fk}. Galaxies with larger dust attenuations tend to have a redder $\beta$ than those with smaller dust attenuations. \citet{Reddy:2010lr} found that typical ($L^*$) star-forming galaxies at $z\sim2-3$ also follow the Meurer relation. However, ultra-luminous infrared galaxies (ULIRGs) and sub-millimeter galaxies (SMGs) with $L_{\rm FIR} >  10^{12}~L_{\sun}$ have larger $L_{\rm FIR}/L_{\rm UV}$ than that inferred from the Meurer relation due to the fact that their UV emission has been highly obscured by dust, and thus the UV-continuum slope does not represent the dust attenuation of the whole galaxy \citep[e.g.,][]{Papovich:2006fk,Howell:2010aa,Penner:2012aa}. 

We measure the FIR luminosity of the local analogs using the Wide-field Infrared Survey Explorer \citep[{WISE;}][]{Wright:2010aa} 12$\mu$m-band flux. We cross-correlate the positions of local analogs with the {\it WISE} DR4 catalog. A total of 196 out of 252 galaxies are detected at 3$\sigma$ in the {\it WISE} 12$\mu$m band, and 50 of the galaxies have UV-continuum slope measurements in section~\ref{sec:uv_slope}. We convert the WISE 12$\mu$m band luminosities to FIR luminosities by adopting the relation between the WISE 12$\mu$m luminosity and the SFR \citep{Lee:2013ab} and the relation between the FIR luminosity and the SFR \citep{Kennicutt:1998aa}. We derive the UV luminosities with the {\it GALEX} FUV magnitudes, which have been corrected for Galactic dust extinction. Figure~\ref{fig_beta_fir} shows the dust attenuation ($L_{\rm FIR}/L_{\rm UV}$) versus the UV-continuum slope ($\beta$) of the local analogs (black filled circles). We also derived the weighted mean value for $L_{\rm FIR}/L_{\rm UV}$ and $\beta$ (red filled circle), which lies on the Meurer relation. Therefore, the local analogs share the same dust attenuation properties with normal star-forming galaxies rather than the highly dust obscured ULIRGs.

\begin{figure}[ht]
\begin{center}
\includegraphics[scale=0.42]{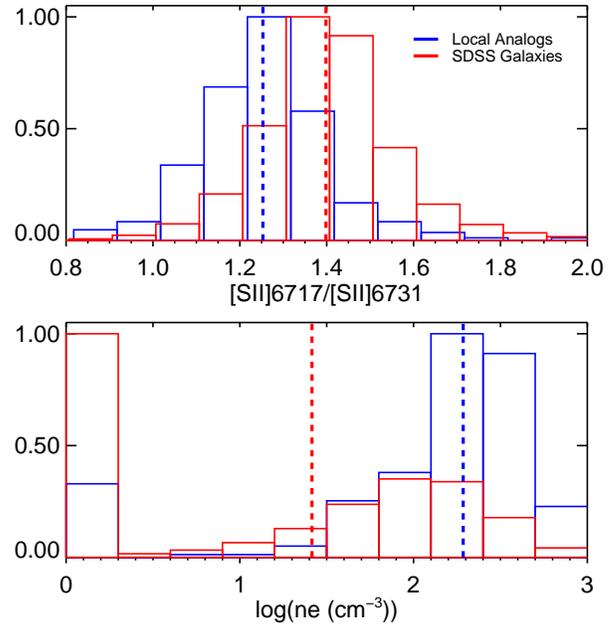}
\caption{Distribution of [SII]$\lambda$6717/[SII]$\lambda$6731 (top panel) and electron density (bottom panel) in the local analogs for high-redshift galaxies (blue histogram) and the SDSS galaxies (red histogram). The blue and red dashed lines represent the median [SII]6717/[SII]6731 and electron density in the local analogs and the SDSS galaxies, respectively.}
\label{fig_ne}
\end{center}
\end{figure}

\subsection{Electron Density}\label{sec:ne}
We compare the electron densities in the local analogs with those in local star-forming galaxies and high-redshift galaxies. High-redshift galaxies often have higher electron densities compared to nearby galaxies. The typical electron densities in nearby star-forming galaxies are less than 100~cm$^{-3}$ \citep[e.g,][]{Zaritsky:1994aa}. A number of studies have suggested that the electron densities in $z\sim2-3$ galaxies can be as high as $n_e\simeq1000$~cm$^{-3}$ based on [\ion{S}{2}]$\lambda\lambda6717,6731$, [\ion{O}{2}]$\lambda\lambda3726,3729$, or \ion{C}{3}]$\lambda\lambda1907,1909$ doublet ratios \citep[e.g.,][]{Hainline:2009aa,Quider:2009lr,Bian:2010vn,Steidel:2014aa}. 

We estimate the electron densities in the local analogs for high-redshift galaxies based on the [\ion{S}{2}]$\lambda$6717/[\ion{S}{2}]$\lambda$6731. The top panel of Figure~\ref{fig_ne} shows the distributions of [\ion{S}{2}]$\lambda$6717/[\ion{S}{2}]$\lambda$6731 in the SDSS galaxies (red histogram) and the local analogs (blue histogram).  The [\ion{S}{2}]$\lambda$6717/[\ion{S}{2}]$\lambda$6731 ratios for a large fraction ($\sim30\%$) of the SDSS galaxies are above the maximum ratio value of 1.43 in the theoretical models, suggesting a low electron density in these galaxies. We assume that the electron density is 1 cm$^{-3}$ in these galaxies. There is an offset between the two [\ion{S}{2}] doublet ratio distributions with a median [\ion{S}{2}] ratio of 1.40 in the local SDSS star-forming galaxies and 1.24 in the local analogs. The lower [\ion{S}{2}] ratio indicates higher electron densities in the analog galaxies. We measured the electron density ($n_e$) using the {\tt nebular.temden} routine in IRAF\citep{Shaw:1995aa}. The bottom panel of Figure~\ref{fig_ne} shows the distribution of the electron densities in the SDSS galaxies (red histogram) and the local analogs (blue histogram). The median electron density in our local analogs is about 200 cm$^{-3}$, which is about an order magnitude higher than than in local star-forming galaxies, and about 10\% of the local analogs have the electron densities greater than $n_e=500$~cm$^{-3}$.

We compare the electron densities in our analogs with those in a sample of star-forming galaxies at $z\sim2.3$ in the MOSDEF survey \citet[]{Sanders:2016aa}. They found that the median electron density at $z\sim2.3$ is about 250~cm$^{-3}$, which is an order of magnitude higher than that in nearby galaxies. This median electron density is about 50 cm$^{-3}$ higher than that in our local analogs. In section~\ref{sec:discussion}, we find that there is an anti-correlation between the electron density and sSFR (see figure~\ref{fig_ssfr_sfrden_ne}). Therefore, the slightly lower electron density in our analogs could be due to the higher sSFRs in our analogs (see section~\ref{sec:sSFR}). 

\begin{figure}[ht]
\begin{center}
\includegraphics[scale=0.36,angle=-90]{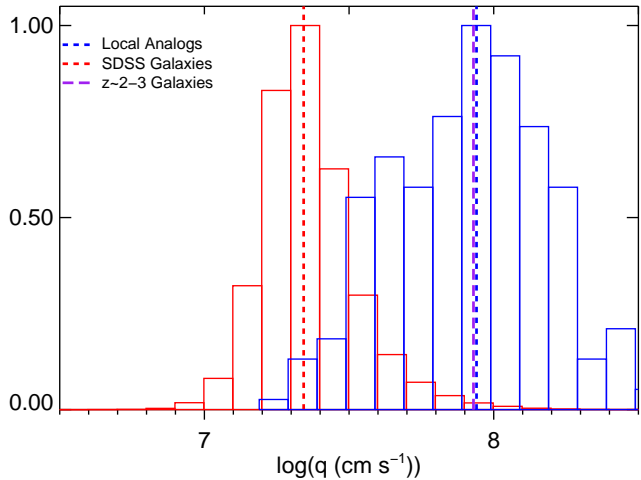}
\caption{Ionization parameter distributions in the local analogs (blue histogram) and SDSS galaxies (red histogram). The red dashed line represents the median ionization parameters in local SDSS galaxies and the blue and purple dashed lines represent the median ionization parameters in the local analogs and high-redshift star-forming galaxies, respectively.}
\label{fig_ion}
\end{center}
\end{figure}

\subsection{Ionization Parameter}\label{sec:q}
The ionization parameters ($q$) in star-forming galaxies at $z\sim2-3$ are higher than those in local star-forming galaxies by about 0.6~dex \citep[e.g.,][]{Kewley:2013ab,Kewley:2013aa,Nakajima:2014aa}. It has been suggested that the high ionization parameter is mainly responsible for the offset between $z\sim2.3$ and local star-forming BPT locus \citep[e.g.,][]{Kewley:2013ab,Kewley:2013aa}. Here, we compare the ionization parameters in the local analogs for high-redshift galaxies with those in low- and high-redshift galaxies. We adopt the same method as used in \citet{Kobulnicky:2004aa} \citep[see also][]{Kewley:2008aa} to measure the ionization parameters in the local analogs. This method is based on the photoionization models from \citet{Kewley:2002fk}. The ionization parameter is determined mainly by using the ratio of [\ion{O}{3}]$\lambda\lambda$4959,5007 and [\ion{O}{2}]$\lambda$3727 (O32). For a fixed O32, the oxygen abundance also affects the ionization parameter measurement. Therefore, the ionization parameter should be corrected for the oxygen abundance derived from the ratio between [\ion{O}{2}]$\lambda$3727+[\ion{O}{3}]$\lambda\lambda$4959,5007 and H$\beta$ (R23). We refer readers to section A2.3 in \citet{Kewley:2008aa} for more details on the ionization parameter measurements.

Figure~\ref{fig_ion} shows the distribution of the ionization parameter in both local analogs for high-redshift galaxies (blue histogram) and the SDSS star-forming galaxies. The local SDSS galaxies have a median $\log(q~{\rm cm~s}^{-1}))=7.29$, which is consistent with the ionization parameters found in nearby star-forming galaxies in other studies \citep[e.g.,][]{Dopita:2006aa,Liang:2006aa,Kewley:2008aa,Nakajima:2014aa}. The ionization parameter in our local analogs is about 0.6~dex higher than that in local star-forming galaxies, with a median ionization parameter of $\log(q~({\rm cm~s}^{-1}))=7.94$ (blue dashed line). This value is in agreement with the ionization parameter found in the $z\sim2-3$ star-forming galaxies, with a median value of $\log(q~({\rm cm~s}^{-1}))=7.91$ \citep[e.g.,][]{Nakajima:2014aa}.

\begin{figure}[h!]
\begin{center}
\includegraphics[scale=0.38,angle=-90]{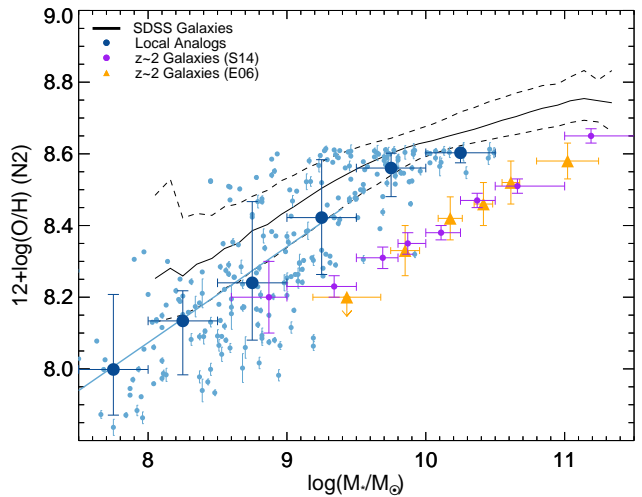}
\caption{Mass-metallicity relation (MZR) of local analogs for high-redshift galaxies, SDSS galaxies, and high-redshift galaxies. The light blue points represent the MZR of individual local analogs, and the dark blue points with error bars represent the median metallicity of local analogs in each stellar mass bin. The horizontal error bar shows the stellar mass range in each mass bin, and the vertical error bar represents the 16th and 84th percentiles of the oxygen abundance distribution in each mass bin. The light blue solid line represents the best linear fit for the local analogs with $M_{*}<10^{9.5}~M_{\sun}$ (Equation~\ref{mzr_fit}). The black solid line represents the median MZR of SDSS galaxies, and two dashed lines represent the 16th and 84th percentiles of the oxygen abundance distribution. The purple points and orange triangles with error bars represents the MZR of $z\sim2.3$ UV-selected star-forming galaxies from \citet[S14]{Steidel:2014aa} and \citet[E06]{Erb:2006rt}, respectively. The horizontal error bar shows the stellar mass range in each mass bin, and the vertical error bar indicates the uncertainty in oxygen abundance estimated from the uncertainty on composite emission-line fluxes for each stellar mass bin \citep{Steidel:2014aa}.}
\label{fig_mz}
\end{center}
\end{figure}

\subsection{Mass-Metallicity Relation}
The relation between galaxy stellar mass and gas-phase metallicities has been established in both low-redshift and high-redshift star-forming galaxies \citep{Tremonti:2004aa,Savaglio:2005aa,Kewley:2008aa,Erb:2006rt,Maiolino:2008lr,Zahid:2013aa,Zahid:2014aa,Steidel:2014aa,Maier:2014aa,Sanders:2015aa}. These studies also suggest a strong evolution of the mass-metallicity relation (MZR) between low- and high-redshift galaxies: high-redshift galaxies have smaller metallicities than galaxies in the local universe at a fixed mass.

We compare the MZRs in the local SDSS galaxies, local analogs, and high-redshift galaxies. We measured the gas-phase oxygen abundance based on the N2 (log([\ion{N}{2}$\lambda6583$]/H$\alpha$) \citep[][PP04 hereafter]{Pettini:2004qe}. In the absence of other strong-line ratios which are sensitive to metallicity, as is often the case in high-redshift galaxies, PP04 calibrated the N2 metallicity diagnostic using $T_e$-based direct metallicity measurements in a sample of local \ion{H}{2} regions. The N2 index has been widely used in high-redshift galaxy metallicity measurements \citep[e.g.,][]{Erb:2006rt,Hainline:2009aa,Bian:2010vn,Steidel:2014aa}. Figure~\ref{fig_mz} shows the mass-metallicity relation in the SDSS star-forming galaxies, the local analogs for high-redshift galaxies, and the $z\sim2-3$ UV-selected star-forming galaxies. The solid line represents the median value of the MZR of the SDSS galaxies, and the light blue filled circles represent the MZR of the local analogs.  We also show the median metallicity of the local analogs in each 0.5 mass bin (dark blue filled circles in Figure~\ref{fig_mz}). For the $z\sim2-3$ galaxies, we show the MZR from \citet[]{Steidel:2014aa}. We fit the MZR of local analogs linearly at the low mass end ($M_{*}<10^{9.5}~M_{\sun}$) because MZR at the high mass end ($M_{*}>10^{9.5}~M_{\sun}$) is significantly affected by our selection methods. The light blue line represents the best log-linear fit with the following equation:
\begin{equation}
12 + \log({\rm O/H}) =8.5 + 0.26\times[\log({M_*}/{M_\sun})-10]; \label{mzr_fit}
\end{equation}
At the low mass end ($M_*=10^{9.0}M_{\sun}$), the MZR of the local analogs is comparable to that at $z\sim2.3$ \citet{Steidel:2014aa}, and is 0.2-0.3~dex lower than the full SDSS sample a fixed stellar mass.

\begin{figure}[t!]
\begin{center}
\includegraphics[scale=0.38,angle=-90]{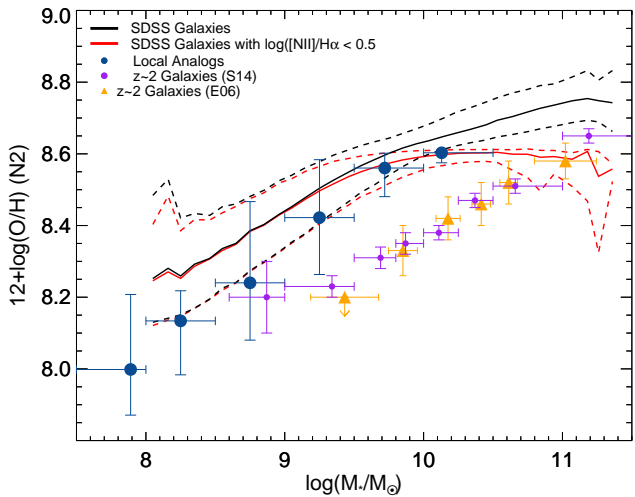}
\caption{Similar to Figure~\ref{fig_mz}, but the red lines represent the MZR of a subsample of SDSS galaxies with log(\ion{N}{2}/H$\alpha$)$ < -0.5$.}
\label{fig_mz_select}
\end{center}
\end{figure}



The properties of the local analogs could be biased by the equation (3) in the selection criterion (log([\ion{N}{2}]/H$\alpha$)$ < -0.5$). This criterion places the upper limit of the metallicity at $12 + \log({\rm O/H}) =8.62$ based on the PP04 calibration, and it also tends to select galaxies with stellar mass less than $5\times10^{10}~M_{\sun}$ due to the MZR. This metallicity cut also reduces the median value of the metallicity for a given stellar mass bin, which could potentially cause the offset of the MZR between SDSS galaxies and local analogs shown in Figure~\ref{fig_mz}.  We apply the same [\ion{N}{2}]/H$\alpha$ cut to the SDSS galaxies to investigate how the line ratio cut affects the MZR. In Figure~\ref{fig_mz_select}, we compare the MZR of the SDSS galaxies (black curve) with that in SDSS galaxies with log([\ion{N}{2}]/H$\alpha$)$ < -0.5$ (red curve). We find that the MZRs are consistent with each other at the low mass end, where $\log(M_{*} (M_\sun))<9.5$. The metallicity in the SDSS galaxies with log(\ion{N}{2}/H$\alpha$)$ < -0.5$ becomes significantly smaller than that in the SDSS galaxies without the metallicity cut when $\log(M_{*} (M_\sun))>9.5$. Therefore, we expect that the selection effect would affect only the MZR of the local analogs at $\log(M_{*} (M_\sun))>9.5$, and that at  $\log(M_{*} (M_\sun))<9.5$, the offset MZR between the analogs and normal SDSS galaxies shown in Figure~\ref{fig_mz} is not due to selection effects.

\section{Comparison with Other Local Analogs}\label{sec:other_analog}
\subsection{Lyman Break Analogs}
\citet{Heckman:2005aa} established a sample of ultraviolet-luminous galaxies (UVLGs) based on their FUV luminosities $L_{\rm FUV} \geq 2\times10^{10}L_{\sun}$. A subset of super compact UVLGs with the highest surface brightnesses ($I_{\rm FUV} \geq 10^9 L_{\sun}/{\rm kpc^2}$) closely resemble the properties of high-redshift Lyman Break Galaxies (LBGs) \citep[e.g.,][]{Hoopes:2007aa}, and thus this type of galaxy is also called Lyman break analogs (LBAs).

We calculate the FUV luminosity of the local analogs selected in this study with 3$\sigma$ detection in the {\it GALEX} FUV band. We find that 35 of our local analogs have a FUV luminosity $L_{\rm FUV} \geq 2\times10^{10}L_{\sun}$, which would have been selected as UVLGs, and 49\% of our local analogs have a FUV surface brightness of $I_{\rm FUV} \geq 10^9 L_{\sun}/{\rm kpc^2}$. The local analogs selected in this study well resemble the FUV/SFR surface density in the LBA and high-redshift LBGs, although they differ in their SFRs and FUV luminosities with the LBAs and LBGs. 

We further study the location in the BPT diagram of the LBAs. We use a sample of UVLGs from \citet{Hoopes:2007aa}, which includes 215 galaxies. We find 35 of them with FUV surface brightness, $I_{\rm FUV} \geq 10^9 L_{\sun}/{\rm kpc^2}$. We obtained the emission line flux of these 35 LBAs by cross-correlating their positions with the MPA-JHU catalog. To make a fair comparison between the LBAs and the local analogs in this work, we apply the same [\ion{N}{2}]/H$\alpha$ cut that has been used for our local analogs ([\ion{N}{2}]/H$\alpha<-0.5$) to the LBAs. We find that 27 LBAs with [\ion{N}{2}]/H$\alpha<-0.5$. Placing the 27 LBAs on the BPT diagram, we find that 22 of them (82\%) are located above the local star-forming sequence defined by equation 3 in \citet{Kewley:2013ab}. This suggests that the LBA also show an offset in the BPT diagram compared to normal local star-forming galaxies. However, the offset is not as large as that has been found in the $z\sim2$ star-forming galaxies. Eight out of the 27 (30\%) LBAs lie within the high-redshift star-forming locus (the region between the two red solid lines in Figure~\ref{fig_select}) in the BPT diagram, and two LBAs (7\%) satisfy our local analog selection criteria (equations 1-3).

We compute the electron densities and ionization parameters in the 27 LBAs. We find a median electron density of $n_e = 110$~cm$^{-3}$ in LBAs, which is about 4 times larger than that in normal SDSS galaxies, but a factor of 2 smaller than that in our local analogs.  \citet{Overzier:2009aa} measured electron density in 29 LBAs using the [\ion{S}{2}] doublet, in which the median [\ion{S}{2}] doublet is 1.3 corresponding to $n_e=132$~cm$^{-3}$. This value is consistent with the median electron density in the LBAs in this study.
Using the O32 and R23 indices, we find a median ionization parameter of $q = 7.78$ in the 27 LBAs, which is smaller than that in our local analogs by $\sim0.15$~dex. Therefore, the local analogs selected in this study more closely resemble the ISM conditions and the ionization radiation field in $z\sim2-3$ star-forming galaxies than the LBAs.

\subsection{Green Pea Galaxies}
The green pea (GP) galaxies are selected based on their strong [\ion{O}{3}] emission line of EW$_0>100$\AA and compact morphologies. \citet{Cardamone:2009aa} selected a sample of green pea galaxies at $0.112<z<0.36$. In this redshift range,  the strong [\ion{O}{3}] emission falls into the SDSS $r$-band filter, separating the GP galaxies from normal galaxies and quasars in the $g-r$ versus $r-i$ color-color diagram \citep[Figure~2 in][]{Cardamone:2009aa}).

We apply the same $g-r$ and $r-i$ color-color selection criteria in \citet{Cardamone:2009aa} to our local analogs. We find that only $\sim20\%$ of the local analogs could be selected as GP galaxies. The majority of our local analogs also have strong [\ion{O}{3}] emission lines: 70\% of the local analogs have a [\ion{O}{3}] rest-frame equivalent width EW$_0>100${\AA}. Most of the galaxies that cannot be selected with the GP method have a lower redshift than the GP redshift selection windows. The [\ion{O}{3}] lines at $z<0.1$ do not fall into the SDSS $r$-band filter, which cannot place our local analogs in the same regions in the $r-i$ version $g-r$ color-color diagram as GP galaxies. We compared the stellar mass, SFR, and sSFRs in the local analogs in this study with the GP galaxies. The median sSFR of the GP galaxies is about 2~Gyr$^{-1}$, which is consistent with that in our local analogs, although the median stellar mass and SFR in our local analogs are 2 times smaller than those in GP galaxies.

We investigate the location of the GP galaxies in the BPT diagram. About 39\% of the GP galaxies satisfy our local analog selection criteria in this study, and 92\% of the GPs galaxies fall into the high-redshift star-forming BPT locus. The median ionization parameter of GP galaxies is 7.93 \citep{Nakajima:2014aa}, which is consistent with that of the local analogs selected in this work.  We compute the electron densities in the GP galaxies and find that the median electron density of the GPs is 160 cm$^{-3}$, which is slightly smaller than that in our local analogs ($n_e\simeq200~\rm cm^{-3}$). These properties suggest that both our local analogs and the GP galaxies have ISM conditions similar to high-redshift galaxies \citep[e.g.,][]{Steidel:2014aa}. Our local analog selection criteria probes galaxies with lower stellar masses and SFRs but with wider ranges of redshift and size distributions. In particular, the selection method used in this study is able to select a significant population ($\sim$50\%) of local analogs for high-redshift galaxies at $z<0.1$.

\subsection{SDSS Galaxies with High Line Luminosity}
\citet{Juneau:2014aa} suggested that the evolution of the ISM conditions from high-redshift ($z\sim1.5$) to low-redshift galaxies may partially be due to the selection effects because only galaxies with high line luminosities can be studied at high-redshift. Therefore, the SDSS galaxies with high line luminosities may also be considered as local analogs for high-redshift galaxies. The line flux limit with signal-to-noise ratio (S/N) of 5 is $4.4\times10^{-18}$~ergs~s$^{-1}$~cm$^{-2}$ in the sample of $z\sim2.3$ star-forming galaxies in \citep{Steidel:2014aa}, which corresponds to an H$\beta$ luminosity of $\log[L_{\rm Line}(\rm ergs~s^{-1})]=41.2$. We select a sample of SDSS galaxies whose luminosities of H$\beta$, H$\alpha$, and [OIII]$\lambda$5007 luminosities are all greater than $\log[L_{\rm Line}(\rm ergs~s^{-1})]=41.2$ and study their locations on the BPT diagram, their ionization parameters, and their electron densities. We find that there is only 6\% of the high line luminosity SDSS galaxies fall into our selection criterion, and 23\% of these SDSS galaxies lie within the high-redshift star-forming locals in the BPT diagram. The median ionization parameter of these SDSS galaxies are 7.64 cm~s$^{-1}$, and the median electron density is 101 cm$^{-3}$. Both are larger than the main sample of SDSS galaxies, but significantly smaller than those in our local analogs. This results suggest that the selection effect on the line luminosity can only partially explain the ISM evolution from high redshift to low redshift. 

\begin{figure*}[ht]
\begin{center}
\includegraphics[scale=0.7,angle=-90]{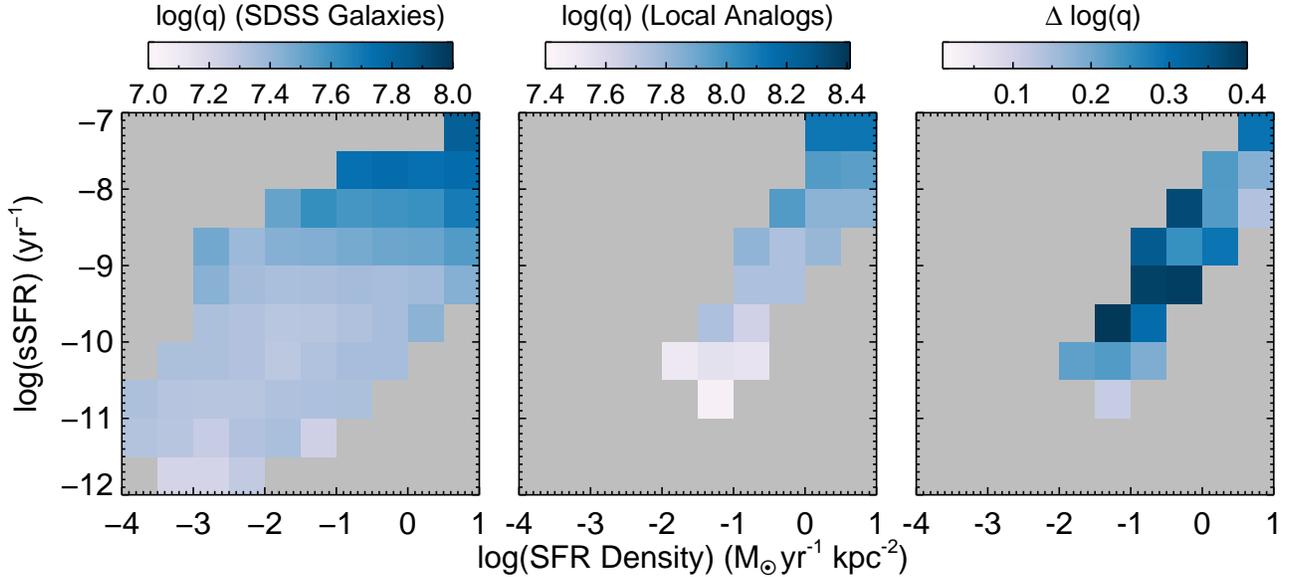}
\caption{Ionization parameter as functions of sSFR and SFR surface density in the SDSS galaxies (left panel), the local analogs in this study (middle panel), and the ionization parameter difference between the local analogs and SDSS galaxies as functions of sSFR and SFR surface density (right panel). The local analogs have higher ionization parameters than the SDSS galaxies for a given sSFR and SFR surface density.}
\label{fig_ssfr_sfrden_q}
\end{center}
\end{figure*}

\begin{figure*}[ht]
\begin{center}
\includegraphics[scale=0.7,angle=-90]{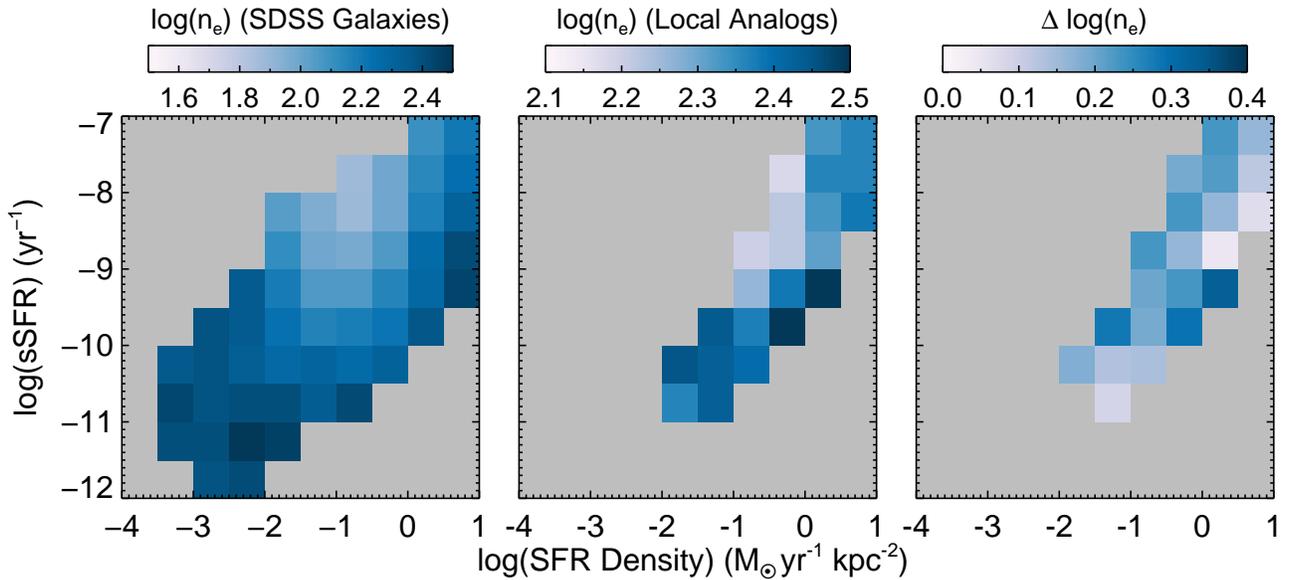}
\caption{Electron density as functions of sSFR and SFR surface density in the SDSS galaxies (left panel), the local analogs in this study (middle panel), and the electron density difference between the local analogs and SDSS galaxies as functions of sSFR and SFR surface density (right panel). The local analogs have higher electron density than the SDSS galaxies for a given sSFR and SFR surface density.}
\label{fig_ssfr_sfrden_ne}
\end{center}
\end{figure*}

\section{Discussion}\label{sec:discussion}
In section~\ref{sec:property}, we have demonstrated that the local analogs have very unique physical ISM conditions, which closely resemble those in high-redshift star-forming galaxies. In this section, we will investigate how the ISM conditions correlate with sSFR and SFR surface density ($\Sigma \rm SFR$), and whether the high ionization parameter and high electron density in our local analogs and high-redshift galaxies are entirely caused by the evolution of sSFR and SFR surface density from high redshift to low redshift.

We use sSFR and SFR surface density to approximately represent the number of ionizing photons per baryon and per volume in a galaxy, respectively. We use the sSFR measurements in section~\ref{sec:sSFR} and compute the SFR surface density by using the following equation, $\Sigma \rm SFR=SFR/(2\pi r_e^2)$, where $r_e$ is the galaxy effective radius from \citet{Blanton:2003aa} in the u-band. We use the optical band to determine the $r_e$ rather than the FUV band because a large fraction of our local analogs cannot be spatially resolved by GALEX FUV images due to their compact size. We divide galaxies into different sSFR and SFR surface density bins and compute the median ionization parameter and electron density in each of the bin to study how the ISM conditions change with the sSFR and SFR surface density. 

Figure~\ref{fig_ssfr_sfrden_q} shows the median ionization parameters in different sSFR and SFR surface density bins. For the SDSS galaxies, there exists a clear trend whereby the ionization parameter increases with increasing sSFR and SFR surface density \citep[see also][]{Brinchmann:2008ab}. We also find a similar trend in our local analogs. We show the difference in ionization between our local analogs and the SDSS galaxies and in each sSFR and SFR surface density bin in the right panel of Figure~\ref{fig_ssfr_sfrden_q}. For a fixed sSFR and SFR surface density, the ionization parameters in the local analogs are systematically higher than those in the SDSS galaxies by 0.25~dex. 

Figure~\ref{fig_ssfr_sfrden_ne} shows the median electron density in difference sSFR and SFR surface density bins. For SDSS galaxies, we find that the electron density increases with increasing SFR surface density for a fixed sSFR, but deceases with increasing sSFR for a fixed SFR surface density. We do not find a similar trend in our local analogs. Due to the relatively large uncertainties on the electron density measurements ($\sim0.3$~dex), a larger sample of local analogs is required to study the relation between electron density and sSFR as well as SFR surface density. The rightmost panel of Figure~\ref{fig_ssfr_sfrden_ne} shows the difference in electron density between our local analogs and the SDSS galaxies in each sSFR and SFR surface density bin. We find that our local analogs have higher electron density than the SDSS galaxies with similar sSFR and SFR surface density by 0.19~dex.

The high sSFR and high SFR surface density can affect the ionization parameter and electron density in a galaxy. However, they can only partially explain the high ionization parameters and high electron densities in the local analogs and high-redshift galaxies. Therefore, additional mechanisms are required to fully explain the evolution of the ISM conditions from the normal nearby galaxies to the local analogs and high-redshift galaxies.

For a ionization-bounded \ion{H}{2} region, the ionization parameter, $q$, can be expressed as follows \citep[e.g.,][]{Charlot:2001aa,Brinchmann:2008aa,Nakajima:2014aa}:
\begin{equation}
q \propto (Qn_{\rm H}\epsilon^2)^\frac{1}{3},
\end{equation}
where $Q$ is the rate of ionizing photons, $n_{\rm H}$ is the number density of hydrogen, and $\epsilon$ is the gas filling factor. We explore how these three parameters influence the ionization parameters as follows.

\begin{itemize}
\item The rate of ionizing photons depends not only on SFR, but also on the hardness of the radiation field. The hardness of the radiation field is sensitive to the slope of the Initial Mass Function (IMF), the star formation history, the metallicity, shocks and AGN activity: (1) A top-heavy IMF (more massive stars for a given total stellar mass) would produce a harder far-UV spectrum. However, there is no clear evidence that the IMF varies over the cosmic time \citep[see][and reference therein]{Bastian:2010aa}. Moreover, \citet{Dopita:2014ac} found that the effect of the IMF slope is rather small, and can only increase  the ionization parameter by 0.1 dex at most; (2) A young stellar population dominated by W-R stars may produce a hard ionizing radiation field. The W-R phase happens at 3-5~Myr after an instantaneous bursts, which could also enhance the ratio of [\ion{O}{3}/H$\beta$] to move star-forming galaxies from the local BPT locus into the high-redshift BPT locus \citep{Brinchmann:2008aa}.  We visually inspect the SDSS spectra of these local analogs to search for the broad component \ion{He}{2}$\lambda$4686 and \ion{N}{3}$\lambda$4640 features, which can be used to identify W-R galaxies \citep{Brinchmann:2008aa}. However, most of the spectra are too noisy to identify these broad components. Further deep optical spectroscopic observations are required to study these W-R features; (3) A harder ionizing field could  be produced by a population of metal-poor stars due to lower stellar atmospheric blanketing and lower mass-loss \citep[e.g.,][]{Levesque:2010aa}. Thus, the ionizing radiation field are likely to be harder in local analogs and high-z galaxies due to their relative lower metallicity; (4) Shocks and AGNs could also also produce a harder radiation field and enhance ionization parameters, and AGNs and shock signatures have been found in high-redshift star-forming galaxies  \citep[e.g.,][]{Newman:2014aa,Genzel:2014ab}. We cross correlate our analogs with the ROSAT ALL-Sky Survey Faint Source Catalog \citep{Voges:2000aa} and find that only one of the targets is detected in the X-ray wavelength. \citet{Greene:2004aa} found that the AGNs hosting low mass black holes ($<2\times10^6~M_\sun$) also show an offset compared to the local star-forming sequence in the BPT diagram. We use the FWHM of H$\alpha$ to understand the possible contamination from broad-line AGNs. We find that the median H$\alpha$ FWHM of our local analogs is about 200~km~s$^{-1}$, which is much smaller than the H$\alpha$ FWHM ($>600$~km~s$^{-1}$) in the broad-line AGNs hosting low mass black holes ($<2\times10^6~M_\sun$) \citep[e.g.][]{Greene:2004aa,Greene:2007aa}. There are only three local analogs in this work with H$\alpha$ FWHM greater than 500~km~s$^{-1}$. These two pieces of evidence suggest that the AGN contamination rate is low in these galaxies.

\item The ionization parameter only mildly depends on the hydrogen density, which can be approached by electron density, $n_e$. We have found that the electron density is 0.19 dex higher in local analogs, which account for a mild increase of about 0.06~dex in ionization parameter, $q$. 

\item The filling factor in the local \ion{H}{2} regions is between 0.01 and 0.1 \citep{Kennicutt:1984aa}. However, this value is very difficult to measure in high-redshift galaxies. 
\end{itemize}

Therefore, it is still unclear which factor is the dominant source causing the high ionization parameter in the our local analogs and high-redshift galaxies in this stage. Further deep long slit and IFS spectroscopic observations will enable us to study the W-R features and obtain spatially resolved line ratios and ionization parameter maps. These follow-up observations will provide clues to solve this problem.

The above discussion is based on the conditions in radiation-bounded clouds. In this case, the size of the \ion{H}{2} region is determined by the ionization equilibrium between the production rate of the ionizing photons and the recombination rate. However, if the surrounding \ion{H}{1} cloud is not enough to absorb all of the ionizing photons, then the ionizing photons from stars will ionize the entire surrounding cloud. In this type of density-bounded cloud, the [\ion{O}{2}] zone becomes smaller, and the [\ion{O}{3}] zone is unlikely to be largely affected, resulting in an enhancement of the [\ion{O}{3}]/[\ion{O}{2}] ratio and an overestimate of the ionization parameters \citep[e.g.,][]{Brinchmann:2008aa,Kewley:2013ab,Nakajima:2014aa}.


\section{Conclusion}
We have presented a sample of local analogs for high-redshift galaxies. These analogs are selected to overlap with the extreme offset galaxies at high redshift in the BPT diagram.  These galaxies are characterized by their ISM and ionization field conditions, which are similar to those in $z\sim2-3$ galaxies. Therefore, this type of local analog is ideal for detailed studies attempting to understand ISM and ionization field conditions in high-redshift galaxies. We summarize the main scientific results as follows:

\begin{enumerate}
\item We use the location of the galaxies in the [\ion{O}{3}]/H$\beta$ versus [\ion{N}{2}]/H$\alpha$ nebular emission-line diagnostic diagram (the BPT diagram) to select local analogs for high-redshift galaxies. These analogs occupy the same locations as the $z\sim2-3$ star-forming galaxies in the BPT diagram. Based on this method, 252 local analogs are selected in a sample of nearby galaxies from the SDSS DR7.

\item We study the properties of the local analogs for high-redshift galaxies. We find that these galaxies share many properties with typical $z\sim2-3$ UV-selected galaxies, including similar high sSFRs, flat UV continuum slopes, dust attenuation properties, and compact morphologies. However, compared to typical $L^*$ UV-selected galaxies at $z\sim2-3$, our analogs have lower FUV luminosity, star formation rate, and stellar mass.

\item Our local analogs have high ionization parameters ($\log q\simeq7.94$~cm$^{-1}$), and high electron densities ($n_e\simeq200$~cm$^{-3}$). The ionization parameter is comparable to that found in the $z\sim2-3$ galaxies and is about 0.6~dex higher than that in the SDSS galaxies. The electron density in our local analogs is an order of magnitude higher than that in the SDSS galaxies. Therefore, this type of local analog well resembles the ionization field and ISM conditions in high-redshift galaxies.

\item We find that the metallicity-mass relation (MZR) in the local analogs is 0.2~dex lower than that in the SDSS galaxies, which is consistent with that found in $z\sim2-3$ UV-selected galaxies \citep{Steidel:2014aa} at the low mass end  ($M_*=10^{9.0}~M_{\sun}$).

\item We compare our local analogs with those selected with other methods, including Lyman break analogs (LBAs) and green pea (GP) galaxies. We find that the UV luminosities in our local analogs are significantly lower than those in the Lyman break analogs. However, the UV surface brightness is comparable to that in the Lyman break analogs and green pea. About 7\% of the Lyman break analogs fall into the BPT region used to select our local analogs, and about 30\% of the Lyman break analogs lie in the BPT region that was defined by $z\sim2$ UV-selected star-forming galaxies \citep{Steidel:2014aa}. We find that both the ionization parameters and the electron density in the Lyman break analogs are significantly smaller than those in our local analogs. The green pea galaxies have a better agreement on the location of the BPT diagram and ionization parameters with the analogs in this study. However, our local analogs have a wider redshift distribution than GP galaxies.

\item We find that the local analogs have higher electron densities and ionization parameters than the SDSS galaxies with similar sSFR and SFR surface density, suggesting that sSFR and SFR surface density are not fully responsible for the difference in the ISM conditions between the local analogs and normal nearby SDSS galaxies. Additional mechanisms are needed to fully explain the difference.

\end{enumerate}

Further studies based on the GAMA, 6dF, and further planned spectroscopic surveys will significantly extend the sample size of the local analogs for high-redshift galaxies. The local analogs selected using this method could provide a great deal of information on the physical conditions of ISM in high-redshift galaxies. Further spatially resolved spectroscopic and high-spatial resolution imaging observations on these galaxies will enable us to resolve the individual star forming regions in these galaxies, study the detailed physical properties in these star-forming regions and make direct comparison with the star-forming regions in high-redshift universe.


\acknowledgments
We would like to thank the anonymous referee for providing constructive comments and help in improving the manuscript. F.B. is grateful to Daniel Stark, Jarle Brinchmann, Kim-Vy Tran, and Andrew Hopkins for their helpful comments and suggestions. We thank the MPA/JHU group for making their catalog public. This research has made use of the VizeR database and Sloan Digital Sky Survey.

{\it Facilities:} \facility{SDSS}


\bibliography{paper}

\begin{thebibliography}{}
\expandafter\ifx\csname natexlab\endcsname\relax\def\natexlab#1{#1}\fi

\bibitem[{{Abazajian} {et~al.}(2009){Abazajian}, {Adelman-McCarthy},
  {Ag{\"u}eros}, {Allam}, {Allende Prieto}, {An}, {Anderson}, {Anderson},
  {Annis}, {Bahcall}, \& et~al.}]{Abazajian:2009aa}
{Abazajian}, K.~N., {Adelman-McCarthy}, J.~K., {Ag{\"u}eros}, M.~A., {et~al.}
  2009, \apjs, 182, 543

\bibitem[{{Andrews} \& {Martini}(2013)}]{Andrews:2013aa}
{Andrews}, B.~H., \& {Martini}, P. 2013, \apj, 765, 140

\bibitem[{{Baldwin} {et~al.}(1981){Baldwin}, {Phillips}, \&
  {Terlevich}}]{Baldwin:1981rr}
{Baldwin}, J.~A., {Phillips}, M.~M., \& {Terlevich}, R. 1981, \pasp, 93, 5

\bibitem[{{Bassett} {et~al.}(2014){Bassett}, {Glazebrook}, {Fisher}, {Green},
  {Wisnioski}, {Obreschkow}, {Mentuch Cooper}, {Abraham}, {Damjanov}, \&
  {McGregor}}]{Bassett:2014aa}
{Bassett}, R., {Glazebrook}, K., {Fisher}, D.~B., {et~al.} 2014, ArXiv
  e-prints, arXiv:1405.6753

\bibitem[{{Bastian} {et~al.}(2010){Bastian}, {Covey}, \&
  {Meyer}}]{Bastian:2010aa}
{Bastian}, N., {Covey}, K.~R., \& {Meyer}, M.~R. 2010, \araa, 48, 339

\bibitem[{{Basu-Zych} {et~al.}(2007){Basu-Zych}, {Schiminovich}, {Johnson},
  {Hoopes}, {Overzier}, {Treyer}, {Heckman}, {Barlow}, {Bianchi}, {Conrow},
  {Donas}, {Forster}, {Friedman}, {Lee}, {Madore}, {Martin}, {Milliard},
  {Morrissey}, {Neff}, {Rich}, {Salim}, {Seibert}, {Small}, {Szalay}, {Wyder},
  \& {Yi}}]{Basu-Zych:2007aa}
{Basu-Zych}, A.~R., {Schiminovich}, D., {Johnson}, B.~D., {et~al.} 2007, \apjs,
  173, 457

\bibitem[{{Basu-Zych} {et~al.}(2009){Basu-Zych}, {Schiminovich}, {Heinis},
  {Overzier}, {Heckman}, {Zamojski}, {Ilbert}, {Koekemoer}, {Barlow},
  {Bianchi}, {Conrow}, {Donas}, {Forster}, {Friedman}, {Lee}, {Madore},
  {Martin}, {Milliard}, {Morrissey}, {Neff}, {Rich}, {Salim}, {Seibert},
  {Small}, {Szalay}, {Wyder}, \& {Yi}}]{Basu-Zych:2009aa}
{Basu-Zych}, A.~R., {Schiminovich}, D., {Heinis}, S., {et~al.} 2009, \apj, 699,
  1307

\bibitem[{{Bian} {et~al.}(2010){Bian}, {Fan}, {Bechtold}, {McGreer}, {Just},
  {Sand}, {Green}, {Thompson}, {Peng}, {Seifert}, {Ageorges}, {Juette},
  {Knierim}, \& {Buschkamp}}]{Bian:2010vn}
{Bian}, F., {Fan}, X., {Bechtold}, J., {et~al.} 2010, \apj, 725, 1877

\bibitem[{{Bian} {et~al.}(2013){Bian}, {Fan}, {Jiang}, {McGreer}, {Dey},
  {Green}, {Maiolino}, {Walter}, {Lee}, \& {Dav{\'e}}}]{Bian:2013aa}
{Bian}, F., {Fan}, X., {Jiang}, L., {et~al.} 2013, \apj, 774, 28

\bibitem[{{Bianchi} {et~al.}(2011){Bianchi}, {Herald}, {Efremova}, {Girardi},
  {Zabot}, {Marigo}, {Conti}, \& {Shiao}}]{Bianchi:2011aa}
{Bianchi}, L., {Herald}, J., {Efremova}, B., {et~al.} 2011, \apss, 335, 161

\bibitem[{{Blanton} {et~al.}(2003){Blanton}, {Hogg}, {Bahcall}, {Baldry},
  {Brinkmann}, {Csabai}, {Eisenstein}, {Fukugita}, {Gunn}, {Ivezi{\'c}},
  {Lamb}, {Lupton}, {Loveday}, {Munn}, {Nichol}, {Okamura}, {Schlegel},
  {Shimasaku}, {Strauss}, {Vogeley}, \& {Weinberg}}]{Blanton:2003aa}
{Blanton}, M.~R., {Hogg}, D.~W., {Bahcall}, N.~A., {et~al.} 2003, \apj, 594,
  186

\bibitem[{{Bouwens} {et~al.}(2007){Bouwens}, {Illingworth}, {Franx}, \&
  {Ford}}]{Bouwens:2007lr}
{Bouwens}, R.~J., {Illingworth}, G.~D., {Franx}, M., \& {Ford}, H. 2007, \apj,
  670, 928

\bibitem[{{Bouwens} {et~al.}(2009){Bouwens}, {Illingworth}, {Franx}, {Chary},
  {Meurer}, {Conselice}, {Ford}, {Giavalisco}, \& {van
  Dokkum}}]{Bouwens:2009uq}
{Bouwens}, R.~J., {Illingworth}, G.~D., {Franx}, M., {et~al.} 2009, \apj, 705,
  936

\bibitem[{{Bouwens} {et~al.}(2012){Bouwens}, {Illingworth}, {Oesch}, {Franx},
  {Labb{\'e}}, {Trenti}, {van Dokkum}, {Carollo}, {Gonz{\'a}lez}, {Smit}, \&
  {Magee}}]{Bouwens:2012fk}
{Bouwens}, R.~J., {Illingworth}, G.~D., {Oesch}, P.~A., {et~al.} 2012, \apj,
  754, 83

\bibitem[{{Brinchmann} {et~al.}(2004){Brinchmann}, {Charlot}, {White},
  {Tremonti}, {Kauffmann}, {Heckman}, \& {Brinkmann}}]{Brinchmann:2004fk}
{Brinchmann}, J., {Charlot}, S., {White}, S.~D.~M., {et~al.} 2004, \mnras, 351,
  1151

\bibitem[{{Brinchmann} {et~al.}(2008{\natexlab{a}}){Brinchmann}, {Kunth}, \&
  {Durret}}]{Brinchmann:2008aa}
{Brinchmann}, J., {Kunth}, D., \& {Durret}, F. 2008{\natexlab{a}}, \aap, 485,
  657

\bibitem[{{Brinchmann} {et~al.}(2008{\natexlab{b}}){Brinchmann}, {Pettini}, \&
  {Charlot}}]{Brinchmann:2008ab}
{Brinchmann}, J., {Pettini}, M., \& {Charlot}, S. 2008{\natexlab{b}}, \mnras,
  385, 769

\bibitem[{{Bruzual} \& {Charlot}(2003)}]{Bruzual:2003lr}
{Bruzual}, G., \& {Charlot}, S. 2003, \mnras, 344, 1000

\bibitem[{{Calzetti} {et~al.}(2000){Calzetti}, {Armus}, {Bohlin}, {Kinney},
  {Koornneef}, \& {Storchi-Bergmann}}]{Calzetti:2000vn}
{Calzetti}, D., {Armus}, L., {Bohlin}, R.~C., {et~al.} 2000, \apj, 533, 682

\bibitem[{{Cardamone} {et~al.}(2009){Cardamone}, {Schawinski}, {Sarzi},
  {Bamford}, {Bennert}, {Urry}, {Lintott}, {Keel}, {Parejko}, {Nichol},
  {Thomas}, {Andreescu}, {Murray}, {Raddick}, {Slosar}, {Szalay}, \&
  {Vandenberg}}]{Cardamone:2009aa}
{Cardamone}, C., {Schawinski}, K., {Sarzi}, M., {et~al.} 2009, \mnras, 399,
  1191

\bibitem[{{Chabrier}(2003)}]{Chabrier:2003wd}
{Chabrier}, G. 2003, \pasp, 115, 763

\bibitem[{{Charlot} \& {Longhetti}(2001)}]{Charlot:2001aa}
{Charlot}, S., \& {Longhetti}, M. 2001, \mnras, 323, 887

\bibitem[{{Daddi} {et~al.}(2004){Daddi}, {Cimatti}, {Renzini}, {Fontana},
  {Mignoli}, {Pozzetti}, {Tozzi}, \& {Zamorani}}]{Daddi:2004fk}
{Daddi}, E., {Cimatti}, A., {Renzini}, A., {et~al.} 2004, \apj, 617, 746

\bibitem[{{Daddi} {et~al.}(2005){Daddi}, {Dickinson}, {Chary}, {Pope},
  {Morrison}, {Alexander}, {Bauer}, {Brandt}, {Giavalisco}, {Ferguson}, {Lee},
  {Lehmer}, {Papovich}, \& {Renzini}}]{Daddi:2005fk}
{Daddi}, E., {Dickinson}, M., {Chary}, R., {et~al.} 2005, \apjl, 631, L13

\bibitem[{{Daddi} {et~al.}(2007){Daddi}, {Dickinson}, {Morrison}, {Chary},
  {Cimatti}, {Elbaz}, {Frayer}, {Renzini}, {Pope}, {Alexander}, {Bauer},
  {Giavalisco}, {Huynh}, {Kurk}, \& {Mignoli}}]{Daddi:2007qy}
{Daddi}, E., {Dickinson}, M., {Morrison}, G., {et~al.} 2007, \apj, 670, 156

\bibitem[{{Daddi} {et~al.}(2010){Daddi}, {Bournaud}, {Walter}, {Dannerbauer},
  {Carilli}, {Dickinson}, {Elbaz}, {Morrison}, {Riechers}, {Onodera}, {Salmi},
  {Krips}, \& {Stern}}]{Daddi:2010fk}
{Daddi}, E., {Bournaud}, F., {Walter}, F., {et~al.} 2010, \apj, 713, 686

\bibitem[{{Dopita} {et~al.}(2014){Dopita}, {Rich}, {Vogt}, {Kewley}, {Ho},
  {Basurah}, {Ali}, \& {Amer}}]{Dopita:2014ac}
{Dopita}, M.~A., {Rich}, J., {Vogt}, F.~P.~A., {et~al.} 2014, \apss, 350, 741

\bibitem[{{Dopita} {et~al.}(2006){Dopita}, {Fischera}, {Sutherland}, {Kewley},
  {Tuffs}, {Popescu}, {van Breugel}, {Groves}, \& {Leitherer}}]{Dopita:2006aa}
{Dopita}, M.~A., {Fischera}, J., {Sutherland}, R.~S., {et~al.} 2006, \apj, 647,
  244

\bibitem[{{Erb} {et~al.}(2006){Erb}, {Shapley}, {Pettini}, {Steidel}, {Reddy},
  \& {Adelberger}}]{Erb:2006rt}
{Erb}, D.~K., {Shapley}, A.~E., {Pettini}, M., {et~al.} 2006, \apj, 644, 813

\bibitem[{{Ferguson} {et~al.}(2004){Ferguson}, {Dickinson}, {Giavalisco},
  {Kretchmer}, {Ravindranath}, {Idzi}, {Taylor}, {Conselice}, {Fall},
  {Gardner}, {Livio}, {Madau}, {Moustakas}, {Papovich}, {Somerville},
  {Spinrad}, \& {Stern}}]{Ferguson:2004lr}
{Ferguson}, H.~C., {Dickinson}, M., {Giavalisco}, M., {et~al.} 2004, \apjl,
  600, L107

\bibitem[{{Fisher} {et~al.}(2014){Fisher}, {Glazebrook}, {Bolatto},
  {Obreschkow}, {Mentuch Cooper}, {Wisnioski}, {Bassett}, {Abraham},
  {Damjanov}, {Green}, \& {McGregor}}]{Fisher:2014aa}
{Fisher}, D.~B., {Glazebrook}, K., {Bolatto}, A., {et~al.} 2014, \apjl, 790,
  L30

\bibitem[{{F{\"o}rster Schreiber} {et~al.}(2009){F{\"o}rster Schreiber},
  {Genzel}, {Bouch{\'e}}, {Cresci}, {Davies}, {Buschkamp}, {Shapiro},
  {Tacconi}, {Hicks}, {Genel}, {Shapley}, {Erb}, {Steidel}, {Lutz},
  {Eisenhauer}, {Gillessen}, {Sternberg}, {Renzini}, {Cimatti}, {Daddi},
  {Kurk}, {Lilly}, {Kong}, {Lehnert}, {Nesvadba}, {Verma}, {McCracken},
  {Arimoto}, {Mignoli}, \& {Onodera}}]{Forster-Schreiber:2009cr}
{F{\"o}rster Schreiber}, N.~M., {Genzel}, R., {Bouch{\'e}}, N., {et~al.} 2009,
  \apj, 706, 1364

\bibitem[{{F{\"o}rster Schreiber} {et~al.}(2011){F{\"o}rster Schreiber},
  {Shapley}, {Genzel}, {Bouch{\'e}}, {Cresci}, {Davies}, {Erb}, {Genel},
  {Lutz}, {Newman}, {Shapiro}, {Steidel}, {Sternberg}, \&
  {Tacconi}}]{Forster-Schreiber:2011qy}
{F{\"o}rster Schreiber}, N.~M., {Shapley}, A.~E., {Genzel}, R., {et~al.} 2011,
  \apj, 739, 45

\bibitem[{{Franx} {et~al.}(2008){Franx}, {van Dokkum}, {Schreiber}, {Wuyts},
  {Labb{\'e}}, \& {Toft}}]{Franx:2008aa}
{Franx}, M., {van Dokkum}, P.~G., {Schreiber}, N.~M.~F., {et~al.} 2008, \apj,
  688, 770

\bibitem[{{Franx} {et~al.}(2003){Franx}, {Labb{\'e}}, {Rudnick}, {van Dokkum},
  {Daddi}, {F{\"o}rster Schreiber}, {Moorwood}, {Rix}, {R{\"o}ttgering}, {van
  der Wel}, {van der Werf}, \& {van Starkenburg}}]{Franx:2003fk}
{Franx}, M., {Labb{\'e}}, I., {Rudnick}, G., {et~al.} 2003, \apjl, 587, L79

\bibitem[{{Genzel} {et~al.}(2010){Genzel}, {Tacconi}, {Gracia-Carpio},
  {Sternberg}, {Cooper}, {Shapiro}, {Bolatto}, {Bouch{\'e}}, {Bournaud},
  {Burkert}, {Combes}, {Comerford}, {Cox}, {Davis}, {Schreiber},
  {Garcia-Burillo}, {Lutz}, {Naab}, {Neri}, {Omont}, {Shapley}, \&
  {Weiner}}]{Genzel:2010aa}
{Genzel}, R., {Tacconi}, L.~J., {Gracia-Carpio}, J., {et~al.} 2010, \mnras,
  407, 2091

\bibitem[{{Genzel} {et~al.}(2011){Genzel}, {Newman}, {Jones}, {F{\"o}rster
  Schreiber}, {Shapiro}, {Genel}, {Lilly}, {Renzini}, {Tacconi}, {Bouch{\'e}},
  {Burkert}, {Cresci}, {Buschkamp}, {Carollo}, {Ceverino}, {Davies}, {Dekel},
  {Eisenhauer}, {Hicks}, {Kurk}, {Lutz}, {Mancini}, {Naab}, {Peng},
  {Sternberg}, {Vergani}, \& {Zamorani}}]{Genzel:2011aa}
{Genzel}, R., {Newman}, S., {Jones}, T., {et~al.} 2011, \apj, 733, 101

\bibitem[{{Genzel} {et~al.}(2013){Genzel}, {Tacconi}, {Kurk}, {Wuyts},
  {Combes}, {Freundlich}, {Bolatto}, {Cooper}, {Neri}, {Nordon}, {Bournaud},
  {Burkert}, {Comerford}, {Cox}, {Davis}, {F{\"o}rster Schreiber},
  {Garc{\'{\i}}a-Burillo}, {Gracia-Carpio}, {Lutz}, {Naab}, {Newman},
  {Saintonge}, {Shapiro Griffin}, {Shapley}, {Sternberg}, \&
  {Weiner}}]{Genzel:2013aa}
{Genzel}, R., {Tacconi}, L.~J., {Kurk}, J., {et~al.} 2013, \apj, 773, 68

\bibitem[{{Genzel} {et~al.}(2014){Genzel}, {F{\"o}rster Schreiber}, {Rosario},
  {Lang}, {Lutz}, {Wisnioski}, {Wuyts}, {Wuyts}, {Bandara}, {Bender}, {Berta},
  {Kurk}, {Mendel}, {Tacconi}, {Wilman}, {Beifiori}, {Brammer}, {Burkert},
  {Buschkamp}, {Chan}, {Carollo}, {Davies}, {Eisenhauer}, {Fabricius},
  {Fossati}, {Kriek}, {Kulkarni}, {Lilly}, {Mancini}, {Momcheva}, {Naab},
  {Nelson}, {Renzini}, {Saglia}, {Sharples}, {Sternberg}, {Tacchella}, \& {van
  Dokkum}}]{Genzel:2014ab}
{Genzel}, R., {F{\"o}rster Schreiber}, N.~M., {Rosario}, D., {et~al.} 2014,
  \apj, 796, 7

\bibitem[{{Gon{\c c}alves} {et~al.}(2014){Gon{\c c}alves}, {Basu-Zych},
  {Overzier}, {P{\'e}rez}, \& {Martin}}]{Goncalves:2014aa}
{Gon{\c c}alves}, T.~S., {Basu-Zych}, A., {Overzier}, R.~A., {P{\'e}rez}, L.,
  \& {Martin}, D.~C. 2014, \mnras, 442, 1429

\bibitem[{{Gon{\c c}alves} {et~al.}(2010){Gon{\c c}alves}, {Basu-Zych},
  {Overzier}, {Martin}, {Law}, {Schiminovich}, {Wyder}, {Mallery}, {Rich}, \&
  {Heckman}}]{Goncalves:2010aa}
{Gon{\c c}alves}, T.~S., {Basu-Zych}, A., {Overzier}, R., {et~al.} 2010, \apj,
  724, 1373

\bibitem[{{Gonz{\'a}lez} {et~al.}(2010){Gonz{\'a}lez}, {Labb{\'e}}, {Bouwens},
  {Illingworth}, {Franx}, {Kriek}, \& {Brammer}}]{Gonzalez:2010fk}
{Gonz{\'a}lez}, V., {Labb{\'e}}, I., {Bouwens}, R.~J., {et~al.} 2010, \apj,
  713, 115

\bibitem[{{Green} {et~al.}(2014){Green}, {Glazebrook}, {McGregor}, {Damjanov},
  {Wisnioski}, {Abraham}, {Colless}, {Sharp}, {Crain}, {Poole}, \&
  {McCarthy}}]{Green:2014aa}
{Green}, A.~W., {Glazebrook}, K., {McGregor}, P.~J., {et~al.} 2014, \mnras,
  437, 1070

\bibitem[{{Greene} \& {Ho}(2004)}]{Greene:2004aa}
{Greene}, J.~E., \& {Ho}, L.~C. 2004, \apj, 610, 722

\bibitem[{{Greene} \& {Ho}(2007)}]{Greene:2007aa}
---. 2007, \apj, 670, 92

\bibitem[{{Hainline} {et~al.}(2009){Hainline}, {Shapley}, {Kornei}, {Pettini},
  {Buckley-Geer}, {Allam}, \& {Tucker}}]{Hainline:2009aa}
{Hainline}, K.~N., {Shapley}, A.~E., {Kornei}, K.~A., {et~al.} 2009, \apj, 701,
  52

\bibitem[{{Heckman} {et~al.}(2005){Heckman}, {Hoopes}, {Seibert}, {Martin},
  {Salim}, {Rich}, {Kauffmann}, {Charlot}, {Barlow}, {Bianchi}, {Byun},
  {Donas}, {Forster}, {Friedman}, {Jelinsky}, {Lee}, {Madore}, {Malina},
  {Milliard}, {Morrissey}, {Neff}, {Schiminovich}, {Siegmund}, {Small},
  {Szalay}, {Welsh}, \& {Wyder}}]{Heckman:2005aa}
{Heckman}, T.~M., {Hoopes}, C.~G., {Seibert}, M., {et~al.} 2005, \apjl, 619,
  L35

\bibitem[{{Heckman} {et~al.}(2011){Heckman}, {Borthakur}, {Overzier},
  {Kauffmann}, {Basu-Zych}, {Leitherer}, {Sembach}, {Martin}, {Rich},
  {Schiminovich}, \& {Seibert}}]{Heckman:2011aa}
{Heckman}, T.~M., {Borthakur}, S., {Overzier}, R., {et~al.} 2011, \apj, 730, 5

\bibitem[{{Hoopes} {et~al.}(2007){Hoopes}, {Heckman}, {Salim}, {Seibert},
  {Tremonti}, {Schiminovich}, {Rich}, {Martin}, {Charlot}, {Kauffmann},
  {Forster}, {Friedman}, {Morrissey}, {Neff}, {Small}, {Wyder}, {Bianchi},
  {Donas}, {Lee}, {Madore}, {Milliard}, {Szalay}, {Welsh}, \&
  {Yi}}]{Hoopes:2007aa}
{Hoopes}, C.~G., {Heckman}, T.~M., {Salim}, S., {et~al.} 2007, \apjs, 173, 441

\bibitem[{{Howell} {et~al.}(2010){Howell}, {Armus}, {Mazzarella}, {Evans},
  {Surace}, {Sanders}, {Petric}, {Appleton}, {Bothun}, {Bridge}, {Chan},
  {Charmandaris}, {Frayer}, {Haan}, {Inami}, {Kim}, {Lord}, {Madore},
  {Melbourne}, {Schulz}, {U}, {Vavilkin}, {Veilleux}, \& {Xu}}]{Howell:2010aa}
{Howell}, J.~H., {Armus}, L., {Mazzarella}, J.~M., {et~al.} 2010, \apj, 715,
  572

\bibitem[{{Hu} {et~al.}(2010){Hu}, {Cowie}, {Barger}, {Capak}, {Kakazu}, \&
  {Trouille}}]{Hu:2010pd}
{Hu}, E.~M., {Cowie}, L.~L., {Barger}, A.~J., {et~al.} 2010, \apj, 725, 394

\bibitem[{{Juneau} {et~al.}(2014){Juneau}, {Bournaud}, {Charlot}, {Daddi},
  {Elbaz}, {Trump}, {Brinchmann}, {Dickinson}, {Duc}, {Gobat}, {Jean-Baptiste},
  {Le Floc'h}, {Lehnert}, {Pacifici}, {Pannella}, \&
  {Schreiber}}]{Juneau:2014aa}
{Juneau}, S., {Bournaud}, F., {Charlot}, S., {et~al.} 2014, \apj, 788, 88

\bibitem[{{Kauffmann} {et~al.}(2003){Kauffmann}, {Heckman}, {White}, {Charlot},
  {Tremonti}, {Brinchmann}, {Bruzual}, {Peng}, {Seibert}, {Bernardi},
  {Blanton}, {Brinkmann}, {Castander}, {Cs{\'a}bai}, {Fukugita}, {Ivezic},
  {Munn}, {Nichol}, {Padmanabhan}, {Thakar}, {Weinberg}, \&
  {York}}]{Kauffmann:2003aa}
{Kauffmann}, G., {Heckman}, T.~M., {White}, S.~D.~M., {et~al.} 2003, \mnras,
  341, 33

\bibitem[{{Kennicutt}(1984)}]{Kennicutt:1984aa}
{Kennicutt}, Jr., R.~C. 1984, \apj, 287, 116

\bibitem[{{Kennicutt}(1998)}]{Kennicutt:1998aa}
---. 1998, \araa, 36, 189

\bibitem[{{Kewley} \& {Dopita}(2002)}]{Kewley:2002fk}
{Kewley}, L.~J., \& {Dopita}, M.~A. 2002, \apjs, 142, 35

\bibitem[{{Kewley} {et~al.}(2013{\natexlab{a}}){Kewley}, {Dopita}, {Leitherer},
  {Dav{\'e}}, {Yuan}, {Allen}, {Groves}, \& {Sutherland}}]{Kewley:2013ab}
{Kewley}, L.~J., {Dopita}, M.~A., {Leitherer}, C., {et~al.} 2013{\natexlab{a}},
  \apj, 774, 100

\bibitem[{{Kewley} {et~al.}(2001){Kewley}, {Dopita}, {Sutherland}, {Heisler},
  \& {Trevena}}]{Kewley:2001aa}
{Kewley}, L.~J., {Dopita}, M.~A., {Sutherland}, R.~S., {Heisler}, C.~A., \&
  {Trevena}, J. 2001, \apj, 556, 121

\bibitem[{{Kewley} \& {Ellison}(2008)}]{Kewley:2008aa}
{Kewley}, L.~J., \& {Ellison}, S.~L. 2008, \apj, 681, 1183

\bibitem[{{Kewley} {et~al.}(2005){Kewley}, {Jansen}, \&
  {Geller}}]{Kewley:2005aa}
{Kewley}, L.~J., {Jansen}, R.~A., \& {Geller}, M.~J. 2005, \pasp, 117, 227

\bibitem[{{Kewley} {et~al.}(2013{\natexlab{b}}){Kewley}, {Maier}, {Yabe},
  {Ohta}, {Akiyama}, {Dopita}, \& {Yuan}}]{Kewley:2013aa}
{Kewley}, L.~J., {Maier}, C., {Yabe}, K., {et~al.} 2013{\natexlab{b}}, \apjl,
  774, L10

\bibitem[{{Kobulnicky} \& {Kewley}(2004)}]{Kobulnicky:2004aa}
{Kobulnicky}, H.~A., \& {Kewley}, L.~J. 2004, \apj, 617, 240

\bibitem[{{Kriek} {et~al.}(2009){Kriek}, {van Dokkum}, {Labb{\'e}}, {Franx},
  {Illingworth}, {Marchesini}, \& {Quadri}}]{Kriek:2009fk}
{Kriek}, M., {van Dokkum}, P.~G., {Labb{\'e}}, I., {et~al.} 2009, \apj, 700,
  221

\bibitem[{{Kroupa}(2001)}]{Kroupa:2001vn}
{Kroupa}, P. 2001, \mnras, 322, 231

\bibitem[{{Lee} {et~al.}(2013){Lee}, {Hwang}, \& {Ko}}]{Lee:2013ab}
{Lee}, J.~C., {Hwang}, H.~S., \& {Ko}, J. 2013, \apj, 774, 62

\bibitem[{{Levesque} {et~al.}(2010){Levesque}, {Kewley}, \&
  {Larson}}]{Levesque:2010aa}
{Levesque}, E.~M., {Kewley}, L.~J., \& {Larson}, K.~L. 2010, \aj, 139, 712

\bibitem[{{Liang} {et~al.}(2006){Liang}, {Yin}, {Hammer}, {Deng}, {Flores}, \&
  {Zhang}}]{Liang:2006aa}
{Liang}, Y.~C., {Yin}, S.~Y., {Hammer}, F., {et~al.} 2006, \apj, 652, 257

\bibitem[{{Liu} {et~al.}(2008){Liu}, {Shapley}, {Coil}, {Brinchmann}, \&
  {Ma}}]{Liu:2008kx}
{Liu}, X., {Shapley}, A.~E., {Coil}, A.~L., {Brinchmann}, J., \& {Ma}, C.-P.
  2008, \apj, 678, 758

\bibitem[{{Maier} {et~al.}(2014){Maier}, {Lilly}, {Ziegler}, {Contini},
  {P{\'e}rez Montero}, {Peng}, \& {Balestra}}]{Maier:2014aa}
{Maier}, C., {Lilly}, S.~J., {Ziegler}, B.~L., {et~al.} 2014, \apj, 792, 3

\bibitem[{{Maiolino} {et~al.}(2008){Maiolino}, {Nagao}, {Grazian}, {Cocchia},
  {Marconi}, {Mannucci}, {Cimatti}, {Pipino}, {Ballero}, {Calura}, {Chiappini},
  {Fontana}, {Granato}, {Matteucci}, {Pastorini}, {Pentericci}, {Risaliti},
  {Salvati}, \& {Silva}}]{Maiolino:2008lr}
{Maiolino}, R., {Nagao}, T., {Grazian}, A., {et~al.} 2008, \aap, 488, 463

\bibitem[{{Masters} {et~al.}(2014){Masters}, {McCarthy}, {Siana}, {Malkan},
  {Mobasher}, {Atek}, {Henry}, {Martin}, {Rafelski}, {Hathi}, {Scarlata},
  {Ross}, {Bunker}, {Blanc}, {Bedregal}, {Dom{\'{\i}}nguez}, {Colbert},
  {Teplitz}, \& {Dressler}}]{Masters:2014aa}
{Masters}, D., {McCarthy}, P., {Siana}, B., {et~al.} 2014, \apj, 785, 153

\bibitem[{{Meurer} {et~al.}(1999){Meurer}, {Heckman}, \&
  {Calzetti}}]{Meurer:1999fk}
{Meurer}, G.~R., {Heckman}, T.~M., \& {Calzetti}, D. 1999, \apj, 521, 64

\bibitem[{{Mosleh} {et~al.}(2011){Mosleh}, {Williams}, {Franx}, \&
  {Kriek}}]{Mosleh:2011aa}
{Mosleh}, M., {Williams}, R.~J., {Franx}, M., \& {Kriek}, M. 2011, \apj, 727, 5

\bibitem[{{Mosleh} {et~al.}(2012){Mosleh}, {Williams}, {Franx}, {Gonzalez},
  {Bouwens}, {Oesch}, {Labbe}, {Illingworth}, \& {Trenti}}]{Mosleh:2012vn}
{Mosleh}, M., {Williams}, R.~J., {Franx}, M., {et~al.} 2012, \apjl, 756, L12

\bibitem[{{Nakajima} \& {Ouchi}(2014)}]{Nakajima:2014aa}
{Nakajima}, K., \& {Ouchi}, M. 2014, \mnras, 442, 900

\bibitem[{{Newman} {et~al.}(2013){Newman}, {Genzel}, {F{\"o}rster Schreiber},
  {Shapiro Griffin}, {Mancini}, {Lilly}, {Renzini}, {Bouch{\'e}}, {Burkert},
  {Buschkamp}, {Carollo}, {Cresci}, {Davies}, {Eisenhauer}, {Genel}, {Hicks},
  {Kurk}, {Lutz}, {Naab}, {Peng}, {Sternberg}, {Tacconi}, {Wuyts}, {Zamorani},
  \& {Vergani}}]{Newman:2013aa}
{Newman}, S.~F., {Genzel}, R., {F{\"o}rster Schreiber}, N.~M., {et~al.} 2013,
  \apj, 767, 104

\bibitem[{{Newman} {et~al.}(2014){Newman}, {Buschkamp}, {Genzel}, {F{\"o}rster
  Schreiber}, {Kurk}, {Sternberg}, {Gnat}, {Rosario}, {Mancini}, {Lilly},
  {Renzini}, {Burkert}, {Carollo}, {Cresci}, {Davies}, {Eisenhauer}, {Genel},
  {Shapiro Griffin}, {Hicks}, {Lutz}, {Naab}, {Peng}, {Tacconi}, {Wuyts},
  {Zamorani}, {Vergani}, \& {Weiner}}]{Newman:2014aa}
{Newman}, S.~F., {Buschkamp}, P., {Genzel}, R., {et~al.} 2014, \apj, 781, 21

\bibitem[{{Noeske} {et~al.}(2007){Noeske}, {Weiner}, {Faber}, {Papovich},
  {Koo}, {Somerville}, {Bundy}, {Conselice}, {Newman}, {Schiminovich}, {Le
  Floc'h}, {Coil}, {Rieke}, {Lotz}, {Primack}, {Barmby}, {Cooper}, {Davis},
  {Ellis}, {Fazio}, {Guhathakurta}, {Huang}, {Kassin}, {Martin}, {Phillips},
  {Rich}, {Small}, {Willmer}, \& {Wilson}}]{Noeske:2007qy}
{Noeske}, K.~G., {Weiner}, B.~J., {Faber}, S.~M., {et~al.} 2007, \apjl, 660,
  L43

\bibitem[{{Overzier} {et~al.}(2010){Overzier}, {Heckman}, {Schiminovich},
  {Basu-Zych}, {Gon{\c c}alves}, {Martin}, \& {Rich}}]{Overzier:2010aa}
{Overzier}, R.~A., {Heckman}, T.~M., {Schiminovich}, D., {et~al.} 2010, \apj,
  710, 979

\bibitem[{{Overzier} {et~al.}(2008){Overzier}, {Heckman}, {Kauffmann},
  {Seibert}, {Rich}, {Basu-Zych}, {Lotz}, {Aloisi}, {Charlot}, {Hoopes},
  {Martin}, {Schiminovich}, \& {Madore}}]{Overzier:2008aa}
{Overzier}, R.~A., {Heckman}, T.~M., {Kauffmann}, G., {et~al.} 2008, \apj, 677,
  37

\bibitem[{{Overzier} {et~al.}(2009){Overzier}, {Heckman}, {Tremonti}, {Armus},
  {Basu-Zych}, {Gon{\c c}alves}, {Rich}, {Martin}, {Ptak}, {Schiminovich},
  {Ford}, {Madore}, \& {Seibert}}]{Overzier:2009aa}
{Overzier}, R.~A., {Heckman}, T.~M., {Tremonti}, C., {et~al.} 2009, \apj, 706,
  203

\bibitem[{{Overzier} {et~al.}(2011){Overzier}, {Heckman}, {Wang}, {Armus},
  {Buat}, {Howell}, {Meurer}, {Seibert}, {Siana}, {Basu-Zych}, {Charlot},
  {Gon{\c c}alves}, {Martin}, {Neill}, {Rich}, {Salim}, \&
  {Schiminovich}}]{Overzier:2011aa}
{Overzier}, R.~A., {Heckman}, T.~M., {Wang}, J., {et~al.} 2011, \apjl, 726, L7

\bibitem[{{Papovich} {et~al.}(2006){Papovich}, {Moustakas}, {Dickinson}, {Le
  Floc'h}, {Rieke}, {Daddi}, {Alexander}, {Bauer}, {Brandt}, {Dahlen}, {Egami},
  {Eisenhardt}, {Elbaz}, {Ferguson}, {Giavalisco}, {Lucas}, {Mobasher},
  {P{\'e}rez-Gonz{\'a}lez}, {Stutz}, {Rieke}, \& {Yan}}]{Papovich:2006fk}
{Papovich}, C., {Moustakas}, L.~A., {Dickinson}, M., {et~al.} 2006, \apj, 640,
  92

\bibitem[{{Penner} {et~al.}(2012){Penner}, {Dickinson}, {Pope}, {Dey},
  {Magnelli}, {Pannella}, {Altieri}, {Aussel}, {Buat}, {Bussmann},
  {Charmandaris}, {Coia}, {Daddi}, {Dannerbauer}, {Elbaz}, {Hwang},
  {Kartaltepe}, {Lin}, {Magdis}, {Morrison}, {Popesso}, {Scott}, \&
  {Valtchanov}}]{Penner:2012aa}
{Penner}, K., {Dickinson}, M., {Pope}, A., {et~al.} 2012, \apj, 759, 28

\bibitem[{{Pettini} \& {Pagel}(2004)}]{Pettini:2004qe}
{Pettini}, M., \& {Pagel}, B.~E.~J. 2004, \mnras, 348, L59

\bibitem[{{Quider} {et~al.}(2009){Quider}, {Pettini}, {Shapley}, \&
  {Steidel}}]{Quider:2009lr}
{Quider}, A.~M., {Pettini}, M., {Shapley}, A.~E., \& {Steidel}, C.~C. 2009,
  \mnras, 398, 1263

\bibitem[{{Reddy} {et~al.}(2012){Reddy}, {Dickinson}, {Elbaz}, {Morrison},
  {Giavalisco}, {Ivison}, {Papovich}, {Scott}, {Buat}, {Burgarella},
  {Charmandaris}, {Daddi}, {Magdis}, {Murphy}, {Altieri}, {Aussel},
  {Dannerbauer}, {Dasyra}, {Hwang}, {Kartaltepe}, {Leiton}, {Magnelli}, \&
  {Popesso}}]{Reddy:2012aa}
{Reddy}, N., {Dickinson}, M., {Elbaz}, D., {et~al.} 2012, \apj, 744, 154

\bibitem[{{Reddy} {et~al.}(2010){Reddy}, {Erb}, {Pettini}, {Steidel}, \&
  {Shapley}}]{Reddy:2010lr}
{Reddy}, N.~A., {Erb}, D.~K., {Pettini}, M., {Steidel}, C.~C., \& {Shapley},
  A.~E. 2010, \apj, 712, 1070

\bibitem[{{Reddy} \& {Steidel}(2009)}]{Reddy:2009kx}
{Reddy}, N.~A., \& {Steidel}, C.~C. 2009, \apj, 692, 778

\bibitem[{{Rodighiero} {et~al.}(2011){Rodighiero}, {Daddi}, {Baronchelli},
  {Cimatti}, {Renzini}, {Aussel}, {Popesso}, {Lutz}, {Andreani}, {Berta},
  {Cava}, {Elbaz}, {Feltre}, {Fontana}, {F{\"o}rster Schreiber},
  {Franceschini}, {Genzel}, {Grazian}, {Gruppioni}, {Ilbert}, {Le Floch},
  {Magdis}, {Magliocchetti}, {Magnelli}, {Maiolino}, {McCracken}, {Nordon},
  {Poglitsch}, {Santini}, {Pozzi}, {Riguccini}, {Tacconi}, {Wuyts}, \&
  {Zamorani}}]{Rodighiero:2011fk}
{Rodighiero}, G., {Daddi}, E., {Baronchelli}, I., {et~al.} 2011, \apjl, 739,
  L40

\bibitem[{{Saintonge} {et~al.}(2013){Saintonge}, {Lutz}, {Genzel}, {Magnelli},
  {Nordon}, {Tacconi}, {Baker}, {Bandara}, {Berta}, {F{\"o}rster Schreiber},
  {Poglitsch}, {Sturm}, {Wuyts}, \& {Wuyts}}]{Saintonge:2013aa}
{Saintonge}, A., {Lutz}, D., {Genzel}, R., {et~al.} 2013, \apj, 778, 2

\bibitem[{{Salmi} {et~al.}(2012){Salmi}, {Daddi}, {Elbaz}, {Sargent},
  {Dickinson}, {Renzini}, {Bethermin}, \& {Le Borgne}}]{Salmi:2012qy}
{Salmi}, F., {Daddi}, E., {Elbaz}, D., {et~al.} 2012, \apjl, 754, L14

\bibitem[{{Salpeter}(1955)}]{Salpeter:1955kx}
{Salpeter}, E.~E. 1955, \apj, 121, 161

\bibitem[{{Sanders} {et~al.}(2015){Sanders}, {Shapley}, {Kriek}, {Reddy},
  {Freeman}, {Coil}, {Siana}, {Mobasher}, {Shivaei}, {Price}, \& {de
  Groot}}]{Sanders:2015aa}
{Sanders}, R.~L., {Shapley}, A.~E., {Kriek}, M., {et~al.} 2015, \apj, 799, 138

\bibitem[{{Sargent} \& {Searle}(1970)}]{Sargent:1970aa}
{Sargent}, W.~L.~W., \& {Searle}, L. 1970, \apjl, 162, L155

\bibitem[{{Savaglio} {et~al.}(2005){Savaglio}, {Glazebrook}, {Le Borgne},
  {Juneau}, {Abraham}, {Chen}, {Crampton}, {McCarthy}, {Carlberg}, {Marzke},
  {Roth}, {J{\o}rgensen}, \& {Murowinski}}]{Savaglio:2005aa}
{Savaglio}, S., {Glazebrook}, K., {Le Borgne}, D., {et~al.} 2005, \apj, 635,
  260

\bibitem[{{Schlegel} {et~al.}(1998){Schlegel}, {Finkbeiner}, \&
  {Davis}}]{Schlegel:1998aa}
{Schlegel}, D.~J., {Finkbeiner}, D.~P., \& {Davis}, M. 1998, \apj, 500, 525

\bibitem[{{Shapley} {et~al.}(2003){Shapley}, {Steidel}, {Pettini}, \&
  {Adelberger}}]{Shapley:2003lr}
{Shapley}, A.~E., {Steidel}, C.~C., {Pettini}, M., \& {Adelberger}, K.~L. 2003,
  \apj, 588, 65

\bibitem[{{Shapley} {et~al.}(2015){Shapley}, {Reddy}, {Kriek}, {Freeman},
  {Sanders}, {Siana}, {Coil}, {Mobasher}, {Shivaei}, {Price}, \& {de
  Groot}}]{Shapley:2015aa}
{Shapley}, A.~E., {Reddy}, N.~A., {Kriek}, M., {et~al.} 2015, \apj, 801, 88

\bibitem[{{Shaw} \& {Dufour}(1995)}]{Shaw:1995aa}
{Shaw}, R.~A., \& {Dufour}, R.~J. 1995, \pasp, 107, 896

\bibitem[{{Shen} {et~al.}(2003){Shen}, {Mo}, {White}, {Blanton}, {Kauffmann},
  {Voges}, {Brinkmann}, \& {Csabai}}]{Shen:2003aa}
{Shen}, S., {Mo}, H.~J., {White}, S.~D.~M., {et~al.} 2003, \mnras, 343, 978

\bibitem[{{Shirazi} {et~al.}(2014){Shirazi}, {Brinchmann}, \&
  {Rahmati}}]{Shirazi:2014ab}
{Shirazi}, M., {Brinchmann}, J., \& {Rahmati}, A. 2014, \apj, 787, 120

\bibitem[{{Spergel} {et~al.}(2007){Spergel}, {Bean}, {Dor{\'e}}, {Nolta},
  {Bennett}, {Dunkley}, {Hinshaw}, {Jarosik}, {Komatsu}, {Page}, {Peiris},
  {Verde}, {Halpern}, {Hill}, {Kogut}, {Limon}, {Meyer}, {Odegard}, {Tucker},
  {Weiland}, {Wollack}, \& {Wright}}]{Spergel:2007fk}
{Spergel}, D.~N., {Bean}, R., {Dor{\'e}}, O., {et~al.} 2007, \apjs, 170, 377

\bibitem[{{Stanway} \& {Davies}(2014)}]{Stanway:2014aa}
{Stanway}, E.~R., \& {Davies}, L.~J.~M. 2014, \mnras, 439, 2474

\bibitem[{{Stark} {et~al.}(2010){Stark}, {Ellis}, {Chiu}, {Ouchi}, \&
  {Bunker}}]{Stark:2010lr}
{Stark}, D.~P., {Ellis}, R.~S., {Chiu}, K., {Ouchi}, M., \& {Bunker}, A. 2010,
  \mnras, 408, 1628

\bibitem[{{Stark} {et~al.}(2013){Stark}, {Schenker}, {Ellis}, {Robertson},
  {McLure}, \& {Dunlop}}]{Stark:2013aa}
{Stark}, D.~P., {Schenker}, M.~A., {Ellis}, R., {et~al.} 2013, \apj, 763, 129

\bibitem[{{Steidel} {et~al.}(2003){Steidel}, {Adelberger}, {Shapley},
  {Pettini}, {Dickinson}, \& {Giavalisco}}]{Steidel:2003kx}
{Steidel}, C.~C., {Adelberger}, K.~L., {Shapley}, A.~E., {et~al.} 2003, \apj,
  592, 728

\bibitem[{{Steidel} {et~al.}(2004){Steidel}, {Shapley}, {Pettini},
  {Adelberger}, {Erb}, {Reddy}, \& {Hunt}}]{Steidel:2004fk}
{Steidel}, C.~C., {Shapley}, A.~E., {Pettini}, M., {et~al.} 2004, \apj, 604,
  534

\bibitem[{{Steidel} {et~al.}(2014){Steidel}, {Rudie}, {Strom}, {Pettini},
  {Reddy}, {Shapley}, {Trainor}, {Erb}, {Turner}, {Konidaris}, {Kulas}, {Mace},
  {Matthews}, \& {McLean}}]{Steidel:2014aa}
{Steidel}, C.~C., {Rudie}, G.~C., {Strom}, A.~L., {et~al.} 2014, \apj, 795, 165

\bibitem[{{Tacconi} {et~al.}(2010){Tacconi}, {Genzel}, {Neri}, {Cox}, {Cooper},
  {Shapiro}, {Bolatto}, {Bouch{\'e}}, {Bournaud}, {Burkert}, {Combes},
  {Comerford}, {Davis}, {Schreiber}, {Garcia-Burillo}, {Gracia-Carpio}, {Lutz},
  {Naab}, {Omont}, {Shapley}, {Sternberg}, \& {Weiner}}]{Tacconi:2010qy}
{Tacconi}, L.~J., {Genzel}, R., {Neri}, R., {et~al.} 2010, \nat, 463, 781

\bibitem[{{Tacconi} {et~al.}(2013){Tacconi}, {Neri}, {Genzel}, {Combes},
  {Bolatto}, {Cooper}, {Wuyts}, {Bournaud}, {Burkert}, {Comerford}, {Cox},
  {Davis}, {F{\"o}rster Schreiber}, {Garc{\'{\i}}a-Burillo}, {Gracia-Carpio},
  {Lutz}, {Naab}, {Newman}, {Omont}, {Saintonge}, {Shapiro Griffin}, {Shapley},
  {Sternberg}, \& {Weiner}}]{Tacconi:2013aa}
{Tacconi}, L.~J., {Neri}, R., {Genzel}, R., {et~al.} 2013, \apj, 768, 74

\bibitem[{{Tremonti} {et~al.}(2004){Tremonti}, {Heckman}, {Kauffmann},
  {Brinchmann}, {Charlot}, {White}, {Seibert}, {Peng}, {Schlegel}, {Uomoto},
  {Fukugita}, \& {Brinkmann}}]{Tremonti:2004aa}
{Tremonti}, C.~A., {Heckman}, T.~M., {Kauffmann}, G., {et~al.} 2004, \apj, 613,
  898

\bibitem[{{Trujillo} {et~al.}(2006){Trujillo}, {F{\"o}rster Schreiber},
  {Rudnick}, {Barden}, {Franx}, {Rix}, {Caldwell}, {McIntosh}, {Toft},
  {H{\"a}ussler}, {Zirm}, {van Dokkum}, {Labb{\'e}}, {Moorwood},
  {R{\"o}ttgering}, {van der Wel}, {van der Werf}, \& {van
  Starkenburg}}]{Trujillo:2006aa}
{Trujillo}, I., {F{\"o}rster Schreiber}, N.~M., {Rudnick}, G., {et~al.} 2006,
  \apj, 650, 18

\bibitem[{{van der Wel} {et~al.}(2014){van der Wel}, {Franx}, {van Dokkum},
  {Skelton}, {Momcheva}, {Whitaker}, {Brammer}, {Bell}, {Rix}, {Wuyts},
  {Ferguson}, {Holden}, {Barro}, {Koekemoer}, {Chang}, {McGrath},
  {H{\"a}ussler}, {Dekel}, {Behroozi}, {Fumagalli}, {Leja}, {Lundgren},
  {Maseda}, {Nelson}, {Wake}, {Patel}, {Labb{\'e}}, {Faber}, {Grogin}, \&
  {Kocevski}}]{van-der-Wel:2014aa}
{van der Wel}, A., {Franx}, M., {van Dokkum}, P.~G., {et~al.} 2014, \apj, 788,
  28

\bibitem[{{Voges} {et~al.}(2000){Voges}, {Aschenbach}, {Boller}, {Brauninger},
  {Briel}, {Burkert}, {Dennerl}, {Englhauser}, {Gruber}, {Haberl}, {Hartner},
  {Hasinger}, {Pfeffermann}, {Pietsch}, {Predehl}, {Schmitt}, {Trumper}, \&
  {Zimmermann}}]{Voges:2000aa}
{Voges}, W., {Aschenbach}, B., {Boller}, T., {et~al.} 2000, VizieR Online Data
  Catalog, 9029, 0

\bibitem[{{Wright} {et~al.}(2010){Wright}, {Eisenhardt}, {Mainzer}, {Ressler},
  {Cutri}, {Jarrett}, {Kirkpatrick}, {Padgett}, {McMillan}, {Skrutskie},
  {Stanford}, {Cohen}, {Walker}, {Mather}, {Leisawitz}, {Gautier}, {McLean},
  {Benford}, {Lonsdale}, {Blain}, {Mendez}, {Irace}, {Duval}, {Liu}, {Royer},
  {Heinrichsen}, {Howard}, {Shannon}, {Kendall}, {Walsh}, {Larsen}, {Cardon},
  {Schick}, {Schwalm}, {Abid}, {Fabinsky}, {Naes}, \& {Tsai}}]{Wright:2010aa}
{Wright}, E.~L., {Eisenhardt}, P.~R.~M., {Mainzer}, A.~K., {et~al.} 2010, \aj,
  140, 1868

\bibitem[{{Wuyts} {et~al.}(2011){Wuyts}, {F{\"o}rster Schreiber}, {van der
  Wel}, {Magnelli}, {Guo}, {Genzel}, {Lutz}, {Aussel}, {Barro}, {Berta},
  {Cava}, {Graci{\'a}-Carpio}, {Hathi}, {Huang}, {Kocevski}, {Koekemoer},
  {Lee}, {Le Floc'h}, {McGrath}, {Nordon}, {Popesso}, {Pozzi}, {Riguccini},
  {Rodighiero}, {Saintonge}, \& {Tacconi}}]{Wuyts:2011fk}
{Wuyts}, S., {F{\"o}rster Schreiber}, N.~M., {van der Wel}, A., {et~al.} 2011,
  \apj, 742, 96

\bibitem[{{York} {et~al.}(2000){York}, {Adelman}, {Anderson}, {Anderson},
  {Annis}, {Bahcall}, {Bakken}, {Barkhouser}, {Bastian}, {Berman}, {Boroski},
  {Bracker}, {Briegel}, {Briggs}, {Brinkmann}, {Brunner}, {Burles}, {Carey},
  {Carr}, {Castander}, {Chen}, {Colestock}, {Connolly}, {Crocker}, {Csabai},
  {Czarapata}, {Davis}, {Doi}, {Dombeck}, {Eisenstein}, {Ellman}, {Elms},
  {Evans}, {Fan}, {Federwitz}, {Fiscelli}, {Friedman}, {Frieman}, {Fukugita},
  {Gillespie}, {Gunn}, {Gurbani}, {de Haas}, {Haldeman}, {Harris}, {Hayes},
  {Heckman}, {Hennessy}, {Hindsley}, {Holm}, {Holmgren}, {Huang}, {Hull},
  {Husby}, {Ichikawa}, {Ichikawa}, {Ivezi{\'c}}, {Kent}, {Kim}, {Kinney},
  {Klaene}, {Kleinman}, {Kleinman}, {Knapp}, {Korienek}, {Kron}, {Kunszt},
  {Lamb}, {Lee}, {Leger}, {Limmongkol}, {Lindenmeyer}, {Long}, {Loomis},
  {Loveday}, {Lucinio}, {Lupton}, {MacKinnon}, {Mannery}, {Mantsch}, {Margon},
  {McGehee}, {McKay}, {Meiksin}, {Merelli}, {Monet}, {Munn}, {Narayanan},
  {Nash}, {Neilsen}, {Neswold}, {Newberg}, {Nichol}, {Nicinski}, {Nonino},
  {Okada}, {Okamura}, {Ostriker}, {Owen}, {Pauls}, {Peoples}, {Peterson},
  {Petravick}, {Pier}, {Pope}, {Pordes}, {Prosapio}, {Rechenmacher}, {Quinn},
  {Richards}, {Richmond}, {Rivetta}, {Rockosi}, {Ruthmansdorfer}, {Sandford},
  {Schlegel}, {Schneider}, {Sekiguchi}, {Sergey}, {Shimasaku}, {Siegmund},
  {Smee}, {Smith}, {Snedden}, {Stone}, {Stoughton}, {Strauss}, {Stubbs},
  {SubbaRao}, {Szalay}, {Szapudi}, {Szokoly}, {Thakar}, {Tremonti}, {Tucker},
  {Uomoto}, {Vanden Berk}, {Vogeley}, {Waddell}, {Wang}, {Watanabe},
  {Weinberg}, {Yanny}, {Yasuda}, \& {SDSS Collaboration}}]{York:2000aa}
{York}, D.~G., {Adelman}, J., {Anderson}, Jr., J.~E., {et~al.} 2000, \aj, 120,
  1579

\bibitem[{{Zahid} {et~al.}(2014){Zahid}, {Dima}, {Kudritzki}, {Kewley},
  {Geller}, {Hwang}, {Silverman}, \& {Kashino}}]{Zahid:2014aa}
{Zahid}, H.~J., {Dima}, G.~I., {Kudritzki}, R.-P., {et~al.} 2014, \apj, 791,
  130

\bibitem[{{Zahid} {et~al.}(2013){Zahid}, {Geller}, {Kewley}, {Hwang},
  {Fabricant}, \& {Kurtz}}]{Zahid:2013aa}
{Zahid}, H.~J., {Geller}, M.~J., {Kewley}, L.~J., {et~al.} 2013, \apjl, 771,
  L19

\bibitem[{{Zaritsky} {et~al.}(1994){Zaritsky}, {Kennicutt}, \&
  {Huchra}}]{Zaritsky:1994aa}
{Zaritsky}, D., {Kennicutt}, Jr., R.~C., \& {Huchra}, J.~P. 1994, \apj, 420, 87

\end{thebibliography}

\end{document}